# Site-selective $d^{10}/d^0$ substitution in a $S$ = ½ spin ladder Ba$_2$CuTe$_{1-x}$W$_x$O$_6$ (0 ≤ $x$ ≤ 0.3)

Charlotte E. Pughe[1], Otto H. J. Mustonen[1*], Alexandra S. Gibbs[2,3,4], Martin Etter[5], Cheng Liu[6], Siân E. Dutton[6], Aidan Friskney[1], Neil C. Hyatt[1], Gavin B. G. Stenning[3], Heather M. Mutch[1], Fiona C. Coomer[7], Edmund J. Cussen[1*]

1. Department of Material Science and Engineering, University of Sheffield, Sheffield S1 3JD, United Kingdom
2. School of Chemistry, University of St Andrews, North Haugh, St Andrews, KY16 9ST
3. ISIS Pulsed Neutron and Muon Source, STFC Rutherford Appleton Laboratory, Didcot OX11 0QX, United Kingdom
4. Max Planck Institute for Solid State Research, Heisenbergstrasse 1, 70569 Stuttgart
5. Deutsches Elektronen-Synchrotron (DESY), 22607 Hamburg, Germany
6. Cavendish Laboratory, University of Cambridge, J.J. Thomson Avenue, Cambridge, CB3 0HE, United Kingdom
7. Johnson Matthey Battery Materials, Reading RG4 9NH, United Kingdom

* Corresponding authors:
Edmund J. Cussen. e.j.cussen@sheffield.ac.uk
Otto H. J. Mustonen. ohj.mustonen@gmail.com

**ABSTRACT**

Remarkably, doping isovalent $d^{10}$ and $d^0$ cations onto the $B''$ site in A$_2$$B'$$B''$O$_6$ double perovskites has the power to direct the magnetic interactions between magnetic $B'$ cations. This is due to changes in orbital hybridization, which favors different superexchange pathways, and leads to the formation of alternative magnetic structures depending on whether $B''$ is $d^{10}$ or $d^0$. Furthermore, the competition generated by introducing mixtures of $d^{10}$ and $d^0$ cations can drive the material into the realms of exotic quantum magnetism. Here, a W$^{6+}$ $d^0$ dopant was introduced to a $d^{10}$ hexagonal perovskite Ba$_2$CuTeO$_6$, which possesses a spin ladder geometry of Cu$^{2+}$ cations, creating a Ba$_2$CuTe$_{1-x}$W$_x$O$_6$ solid solution ($x$ = 0 - 0.3). Neutron and synchrotron X-ray diffraction show that W$^{6+}$ is almost exclusively substituted for Te$^{6+}$ on the corner-sharing site within the spin ladder, in preference to the face-sharing site between ladders. This means the intra-ladder interactions are selectively tuned by the $d^0$ cations. Bulk magnetic measurements suggest this suppresses magnetic ordering in a similar manner to that observed for the spin-liquid like material Sr$_2$CuTe$_{1-x}$W$_x$O$_6$. This further demonstrates the utility of $d^{10}$ and $d^0$ dopants as a tool for tuning magnetic ground states in a wide range of perovskites and perovskite-derived structures.

**1. INTRODUCTION** Chemical doping is widely used to tune, control and influence the properties of materials. The periodic table offers a plethora of dopants to choose from based on differences in charge and ionic radii. By careful selection, it is possible to desirably modify the structural, electronic and magnetic properties without the need for external controls (i.e. temperature, pressure or magnetic field), and in some cases generate entirely different behaviors to the parent compound. The classic example is $Sr^{2+}$ doping of the antiferromagnetic layered perovskite-type $La_2CuO_4$ that leads to high $T_C$ superconductivity in $La_{2-x}Sr_xCuO_4$ ($x$ = 0.06-0.25).[1–3] This discovery has fascinated scientists for decades and led to a cascade of studies investigating low-dimensional copper systems.

Dopants have such dramatic effects because they intrinsically modify the interactions within the parent material. In magnetic oxides, these interactions are typically superexchange interactions mediated by oxygen anions. These interactions are generally well understood when the magnetic cations are connected by a single oxygen anion.[4] However, the situation is more complicated when the magnetic cations are further away and occur by extended superexchange. Recently, a new method to directly tune these extended superexchange interactions has been developed.[5] This method is based on doping diamagnetic $d^{10}$ and $d^0$ cations into extended superexchange pathways that link magnetic cations. This $d^{10}/d^0$ effect can be used in $A_2B'B''O_6$ double perovskites, where $B'$ is a magnetic cation and $B''$ is a diamagnetic $d^{10}$ or $d^0$ cation.[6,7] The double perovskite structure consists of corner-sharing $B'O_6$ and $B''O_6$ octahedrons alternating in a rock salt-type order (Figure 1a).[8] The superexchange between the magnetic $B'$ cations is extended via orbital overlap with the linking $B''$ cations and O $2p$ orbitals (i.e. $B'$-O-$B''$-O-$B'$).

We have recently shown that diamagnetic $d^{10}$ and $d^0$ cations on the linking -$B''$- site have a significant effect on the magnetic interactions and ground states in double perovskites.[6,9] We investigated this $d^{10}/d^0$ effect in the cubic double perovskites $Ba_2MnTeO_6$ and $Ba_2MnWO_6$, in which the magnetic $Mn^{2+}$ cations are linked by either $4d^{10}$ $Te^{6+}$ or $5d^0$ $W^{6+}$ cations. In these isostructural materials, $Mn^{2+}$ $S$ = 5/2 magnetism is described using a simple fcc Heisenberg model consisting of a 90° (nearest neighbor (NN) - $J_1$) and 180° (next-nearest neighbor (NNN) - $J_2$) Mn-O-(Te/W)-O-Mn interaction (Figure 1a). Neutron scattering experiments demonstrated the dominant interaction heavily depends on the nonmagnetic $B''$ cation, with a stronger $J_1$ when $B''$ = $Te^{6+}$ ($4d^{10}$) and a stronger $J_2$ when $B''$ = $W^{6+}$ ($5d^0$). The contrasting $J_1$ and $J_2$ interactions produce entirely different magnetic structures for $Ba_2MnTeO_6$ (type I-AFM) and $Ba_2MnWO_6$ (type II-AFM). The $d^{10}/d^0$ effect arises as the filled $Te^{6+}$ $4d^{10}$ orbitals hybridize weakly with O $2p$ compared to empty $W^{6+}$ $5d^0$.[10] This limits $Te^{6+}$ participation in $J_2$ exchange, where a $d$-orbital contribution from the $B''$ cation is required. We also highlighted that the $d^{10}/d^0$ effect extends beyond simple cubic structures to a large range of $3d$ transition metal $B'$ = Co[11,12], Ni[13–15] and Cu[16–19] double perovskites, all of which follow the same principle based on the nonmagnetic $B''$-site: $d^0$ - strong $J_2$ (type II) or $d^{10}$ - strong $J_1$ (type I/Néel order).

The most striking example of the $d^{10}/d^0$ effect in 3$d$ double perovskites are the $Cu^{2+}$ $S$ = 1/2 compounds $Sr_2CuTeO_6$, $Sr_2CuWO_6$ and their solid solution $Sr_2CuTe_{1-x}W_xO_6$, where the $d^{10}/d^0$ doping stabilizes a novel quantum disordered ground state. Here, the combination of the $Cu^{2+}$ Jahn-Teller (J-T) effect and orbital ordering produces a square-lattice Heisenberg antiferromagnet, with highly two-dimensional magnetism.[17,20,21] The tetragonal unit cell has square lattice $ab$-planes of $Cu^{2+}$ cations in which superexchange is described using in-plane $J_1$ (NN) and $J_2$ (NNN) interactions, but with additional weak inter-plane interactions ($J_3$ and $J_4$) along $c$ (Figure 1b).[17,18] Following the principles of the $d^{10}/d^0$ effect, $Sr_2CuTeO_6$ is Néel ordered, while a strong $J_2$ leads to columnar ordering for $Sr_2CuWO_6$.[16,19,21–24] Upon introducing a mixture of $W^{6+}$ and $Te^{6+}$, the material evades ordering over a wide portion ($x$ = 0.05 - 0.6) of the $Sr_2CuTe_{1-x}W_xO_6$ solution.[7,10,25–28] The 50:50 mixture $Sr_2CuTe_{0.5}W_{0.5}O_6$ closely resembles a quantum spin-liquid, an exotic magnetic state where the moments remain dynamic at zero Kelvin and have been highly sought since they were first proposed in the 1970s.[29–32]

The question remains whether similarly exotic magnetic behavior can be introduced by $d^{10}/d^0$ doping in other magnetic lattices than the square-lattice, and whether this can be extended from perovskites to perovskite-derived structures? Depending on the choice of $A$ and $B'/B''$ cations, $B'$-O-$B''$-O-$B'$ linkers form between corner-sharing or/and face-sharing octahedra, generating the classic double perovskite structure in the purely corner-sharing case, while the introduction of face-sharing leads to the hexagonal perovskite structure.[33–35] The observation of similarly suppressed magnetism in structures with different octahedral connectivity would suggest competing $d^{10}$ vs $d^0$ interactions can be employed in a range of materials to access novel quantum states, many of which are hard to realize experimentally.[36] $Ba_2CuTeO_6$ is an excellent system to test this due to its hexagonal perovskite structure that results in a spin ladder magnetic geometry.[37–40] Within the spin ladder, $Cu^{2+}$ cations are linked via three key Cu-O-Te-O-Cu exchange interactions illustrated in Figure 1c. These are the intra-ladder $J_{leg}$ and $J_{rung}$ interactions via the corner-sharing $Te(1)O_6$ units and the inter-ladder interaction via the face-sharing $Te(2)O_6$ units within the Cu-Te(2)-Cu trimers.[41] Given there are two $B''$ sites onto which $W^{6+}$ can be doped, this offers the possibility of tuning the $J_{leg}$ and $J_{rung}$ interactions independently of the $J_{inter}$ i.e. a more complex phase space than cubic perovskites. For clarity the two $B''$ sites are henceforth labelled $B''$(c) and $B''$(f), where c and f denote corner and face-sharing. As shown in Figure 1d, the intra-ladder interactions in $Ba_2CuTeO_6$ bear close similarity to the Cu-O-Te-O-Cu interactions within the square-lattice of $Sr_2CuTe_{1-x}W_xO_6$ i.e. four corner $Cu^{2+}$ cations interacting via $B'$-O-$B''$-O-$B'$ superexchange. In addition, the significant $J_{inter}$ leads to formation of a Néel ordered ground state for Ba$_2$CuTeO$_6$, the same type of ordering observed for Sr$_2$CuTeO$_6$.[38] Hence, in an analogous manner to Sr$_2$CuTe$_{1-x}$W$_x$O$_6$, one might expect similar strong suppression of magnetic order upon site-specific doping of $B'' = d^0$ cations onto the intra-ladder $B''$(c) sites in Ba$_2$CuTe$_{1-x}$W$_x$O$_6$.[25,27,28]

To answer these questions, we prepared the Ba$_2$CuTe$_{1-x}$W$_x$O$_6$ solid solution (0 ≤ $x$ ≤ 0.3). Using a combination of crystallographic and spectroscopic techniques, we show that W$^{6+}$ can be site-selectively doped onto the corner-sharing $B''$(c) site in Ba$_2$CuTe$_{1-x}$W$_x$O$_6$. Bulk magnetic characterization demonstrates this has a major effect on the Néel ordered state, suggesting formation of a quantum disordered ground state similar to Sr$_2$CuTe$_{1-x}$W$_x$O$_6$. This demonstrates the $d^{10}/d^0$ effect can be extended to perovskite-derived structures such as hexagonal perovskites.

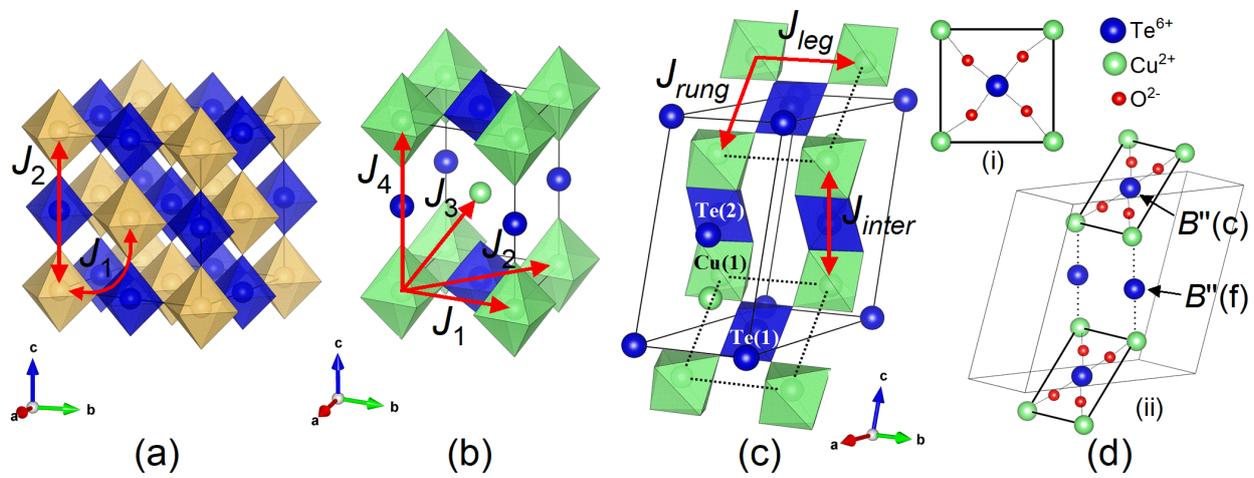

**Figure 1.** Magnetic interactions in $B'' = $ W$^{6+}$ ($d^0$) and/or Te$^{6+}$ ($d^{10}$) perovskite structures: (a) The simple fcc Heisenberg $J_1$ and $J_2$ interactions in the cubic double perovskites Ba$_2$Mn(Te/W)O$_6$; (b) Heisenberg square-lattice interactions in the Sr$_2$Cu(Te/W)O$_6$, Ba$_2$CuWO$_6$ and Ba$_2$CuTeO$_6$ high pressure tetragonal perovskites; (c) Spin ladder interactions in the hexagonal perovskite Ba$_2$CuTeO$_6$; and (d) Extended superexchange within the square lattice planes of Sr$_2$CuTeO$_6$ (i) and spin ladders in Ba$_2$CuTeO$_6$ (ii). Note, the similarity between the Cu-O-Te-O-Cu exchange pathways. The corner-sharing $B''$(c) and face-sharing $B''$(f) Te sites in Ba$_2$CuTeO$_6$ are indicated by the arrows in d(ii).

## 2. EXPERIMENTAL

**2.1 SYNTHESIS** Conventional solid-state chemistry techniques were used to synthesize polycrystalline samples of $Ba_2CuTe_{1-x}W_xO_6$. Compositions $x$ = 0, 0.05, 0.1, 0.2 and 0.3 were prepared by thoroughly mixing stochiometric quantities of high purity $BaCO_3$ (99.997%), CuO (99.9995%), $TeO_2$ (99.995%) and $WO_3$ (99.998%) (all purchased from Alfa Aesar) in an agate mortar. The reactant mixtures were pelletized and calcined at 900 °C in air, before being fired at 1000-1100 °C for 24-hour periods with intermittent grinding. Phase purity was monitored using X-ray diffraction (Rigaku Miniflex, Cu $K\alpha$). A total of 72 – 120 hours was required to achieve phase purity in all compositions, with the heating time increasing as the W content increased. The synthesis was stopped once single-phase samples were obtained.

**2.2 MAGNETIZATION AND HEAT CAPACITY MEASUREMENTS** Magnetic characterization was performed using a Quantum Design MPMS3 magnetometer. ~100 mg of powder was sealed in a gelatin capsule, that was then secured in a polymer straw sample holder. Zero-field cooled (ZFC) and field cooled (FC) curves were measured between 2-300 K in DC SQUID mode using an external field of 0.1 T. Heat capacity measurements were performed using a Quantum Design PPMS instrument. The samples were mixed with silver (99.999%) in a 1:1 gravimetric ratio to enhance the low temperature thermal conductivity. The $Ba_2CuTe_{1-x}W_xO_6$:Ag powder was pressed into a pellet. The pellet was broken and shards weighing ~10 mg were selected, and the heat capacity measured between 2-120 K using the thermal relaxation method. The silver contribution was removed based on a measurement of pure silver powder.

**2.3 NEUTRON POWDER DIFFRACTION** The nuclear structure of $x$ = 0.05, 0.1 and 0.3 was investigated using the High-Resolution Powder Diffractometer (HRPD) at the ISIS Neutron and Muon Source. ~8g of each sample powder was loaded into an Al-alloy slab-can and sealed using vanadium windows. The exposed surfaces of the slab-can were shielded using highly absorbing Gd and Cd foils so that only the vanadium windows of the can were exposed to the neutron flux. After aligning the slab-can perpendicular to the neutron beam, time-of-flight neutron powder diffraction patterns were recorded between 2 – 300 K using a cryostat to cool the sample. The data can be found online.[42] The data were corrected for sample absorption and Rietveld refinements were performed using GSAS-2.[43,44] VESTA was used to visualize the crystal structures.[45]

**2.4 SYNCHROTRON X-RAY DIFFRACTION** Samples of $x$ = 0.1, 0.2 and 0.3 were loaded into glass capillaries 0.6 mm in diameter and measured at room temperature on the P02.1 beamline at the PETRA III X-ray radiation source (DESY) using a wavelength of $\lambda$ = 0.20742 Å. The capillary was located 1169.45 mm from the Perkin Elmer XRD1621 2D detector and spun during the measurement. The 2D data were processed using DAWN Science. The 1D data obtained were refined using GSAS-2. Both the X-ray and neutron data (as well as the bulk magnetization data) were collected using samples from the same batch.

**2.5 EXTENDED X-RAY ABSORPTION FINE STRUCTURE (EXAFS)** EXAFS measurements were performed on the Beamline for Materials Measurement (6-BM) beamline at the National Synchrotron Light Source II (NSLS-II). Room temperature X-ray absorption spectra (XAS) of $x$ = 0.3 were recorded in transmission mode near the W $L_3$ edge, using a finely ground specimen dispersed in polyethylene glycol to achieve a thickness of one absorption length. Incident and transmitted beam intensities were measured using ionization chambers, filled with mixtures of He and $N_2$, operated in a stable region of their *I/V* curve. A tungsten foil was used as an internal energy calibration where the first inflection point in the measured W $L_3$ edge was defined to be $E_0$ = 10206.8 eV. Data reduction and analysis was performed using the programmes Athena, Artemis and Hephaestus.[46] The EXAFS data was then analyzed using ATOMS and FEFF in the Artemis package with the monoclinic structural model.

## 3. RESULTS

**3.1 CRYSTAL STRUCTURE** Initial structural characterization was achieved using laboratory X-ray diffraction. The laboratory X-ray diffraction patterns (XRD) confirm phase purity between $x$ = 0 to 0.3, after which there are significant $W^{6+}$ impurities that are not diminished even after further heating. Hence, the $x$ = 0.3 composition lies close to the solubility limit. Rietveld refinement showed all the $Ba_2CuTe_{1-x}W_xO_6$ structures adopt the same $C2/m$ symmetry as the $Ba_2CuTeO_6$ parent structure. The unit cell volume decreases linearly with $x$ (see inset in supplementary Figure S1); this showing Vegard's law behavior indicates successful $W^{6+}$ doping into the $Ba_2CuTeO_6$ structure. Synchrotron X-ray diffraction, Extended X-ray Absorption Fine Structure (EXAFS) and neutron diffraction studies provided further insight into the structural effects of doping across the solution.

**3.1.1 Synchrotron X-ray diffraction** Figure2a shows an illustrative synchrotron X-ray diffraction pattern collected for the $x$ = 0.2 sample. Data collected at 300 K for $x$ = 0.1, 0.2 and 0.3 were all used to test three possible site occupancy models. These models are: (1) $W^{6+}$ exclusively on the *B''*(c) site; (2) $W^{6+}$ exclusively on the *B''*(f) site; and (3) $W^{6+}$ occupies both *B''*(c) and *B''*(f) sites. An equal distribution (50:50) was initially assumed in model (3). The results in Figure 2a and 2b show model (1) reproduces the observed diffraction profile uniquely well. Figure 2b shows $R_{wp}$ is consistently lower when $W^{6+}$ exclusively occupies the *B''*(c) site in $x$ = 0.1, 0.2 and 0.3. This suggests a strong preference for corner-sharing, which was further evaluated by allowing the site-occupancies to refine, within the constraints of sample stoichiometry. This identified a small amount of $W^{6+}$ on the *B''*(f) site in each sample, with the site occupancy increasing linearly with $x$ up to a maximum value of *ca.* 3% as shown in Table 1. In each refinement 5% of the $W^{6+}$ present in the sample is found on the face-sharing octahedral site. Comparing the $R_{wp}$ and $\chi^2$ values in Table 1 to those in Figure 2(b) shows minor occupation of the *B''*(f) site by $W^{6+}$ leads to a slight, but not negligible improvement in the fit compared to model (1). $W^{6+}$ occupancy of the *B''*(f) site was confirmed by refinements using the initial site occupancies from model (1) and model (2) as starting values. Both refinements converge to the same results in Table 1. Given the energetics for ion migration diminishes on cooling from room temperature, the site preference undoubtedly extends to the low temperature structures.

**Table 1.** Refined *B''*(c) and *B''*(f) site fractions determined using the $x$ = 0.1, 0.2 and 0.3 synchrotron X-ray diffraction data. The W(1) and W(2) site fractions were used to calculate the percentage of the total amount of $W^{6+}$ on the *B''*(f) site in each composition. Also shown are the $R_{wp}$ and $\chi^2$ values for the Rietveld fits.

|  | *B''*(c) | | *B''*(f) | | Percentage of total $W^{6+}$ on *B''*(f) site | $R_{wp}$ | $\chi^2$ |
|---|---|---|---|---|---|---|---|
|  | Te(1) | W(1) | Te(2) | W(2) |  |  |  |
| $x$ = 0.1 | 0.809(1) | 0.191(1) | 0.991(1) | 0.009(1) | 4.7(2)% | 1.54 | 3.39 |
| $x$ = 0.2 | 0.618(1) | 0.382(1) | 0.982(1) | 0.018(1) | 4.7(2)% | 1.75 | 4.54 |
| $x$ = 0.3 | 0.430(1) | 0.570(1) | 0.970(1) | 0.030(1) | 5.3(2)% | 2.60 | 10.50* |

* The larger $\chi^2$ for $x$ = 0.3 reflects a longer counting time compared to the $x$ = 0.1 and 0.2 samples.

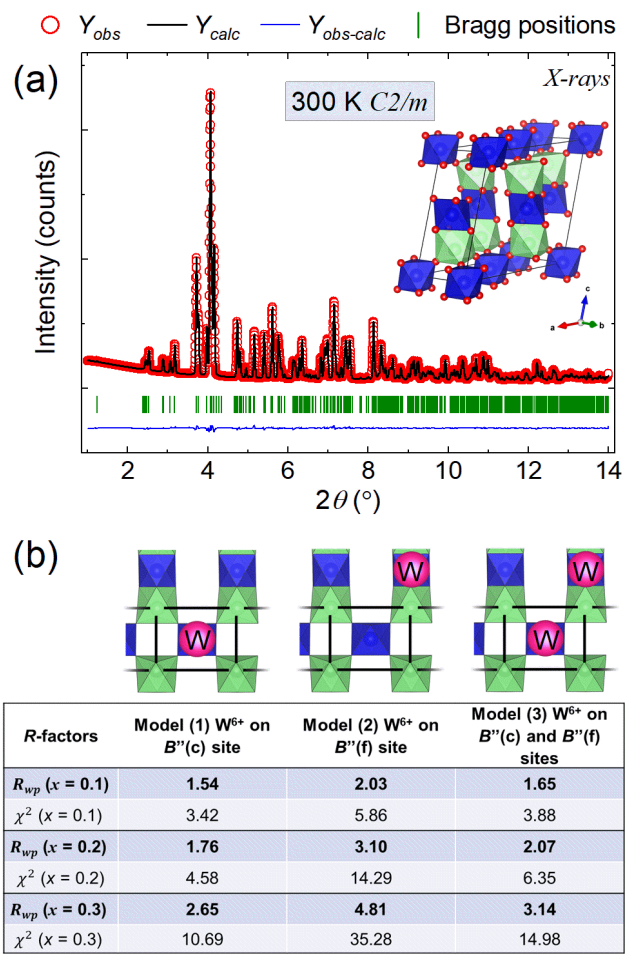

**Figure 2.** (a) Synchrotron X-ray diffraction pattern of $Ba_2CuTe_{0.8}W_{0.2}O_6$ at room temperature collected using a wavelength of $\lambda$ = 0.20742 Å and (b) R-factors obtained from Rietveld refinement using the three different $W^{6+}$ site occupancy models for $Ba_2CuTe_{1-x}W_xO_6$. The crystal structures directly above the R-factors for each model depict the placement of $W^{6+}$ (shown in pink) on either the corner-sharing B''(c) site, face-sharing B''(f) site, or both the B''(c) and B''(f) sites (50:50) in the x = 0.1, 0.2 and 0.3 compositions. The $Te^{6+}$ are colored blue and the $Cu^{2+}$ cations in the spin ladder green.

**3.1.2 Extended X-ray Absorption Fine Structure (EXAFS)** Analysis of W $L_3$ EXAFS data considered the following models: (1) full $W^{6+}$ substitution on the B''(c) site; and (2) full $W^{6+}$ substitution on the B''(f) site, within the monoclinic structure. The methodology is further elaborated in the supplementary information. Model (1) afforded a plausible $W^{6+}$ environment at the B''(c) site, with reasonable path lengths and positive Debye-Waller factors (Table S11). Model (1) presented an excellent fit to the data as evidenced by the graphical fits in Figure 3a, with an R-factor of 1.18%. In contrast, model (2) did not afford a plausible $W^{6+}$ environment at the B''(f) site, with several paths having negative Debye-Waller factors (Table S12). The graphical fits for model (2) in Figure 3b show obvious regions of poor fit evident when compared to model (1) in Figure 3a, and produces a comparatively higher $R$-factor of 9.07%. On inspection, it was evident that $W^{6+}$ substitution on the $B''$(f) site does not provide adequate scattering paths to fit the significant second shell contribution observed in the $\chi(R)$ transform of the EXAFS data in the range $3 < R < 4$ Å (compare Figures S16 and S18). Furthermore, attempting to model $W^{6+}$ substitution on the Cu site proved fruitless, with the refined model evidencing implausible path lengths, negative Debye-Waller factors, and a negative passive electron reduction factor ($S_0^2$). This analysis provides compelling evidence for dominant substitution of $W^{6+}$ at the $B''$(c) site, in agreement with other data presented herein, but has the advantage of providing an element specific perspective. Attempts were made to fit the EXAFS data using contributions from both models (1) and (2), under a linear constraint, to assess the potential for disorder of a fraction of $W^{6+}$ from the $B''$(c) to $B''$(f) site. However, it was not possible to adequately stabilize such a fit, since the number of variables approached the number of data points.

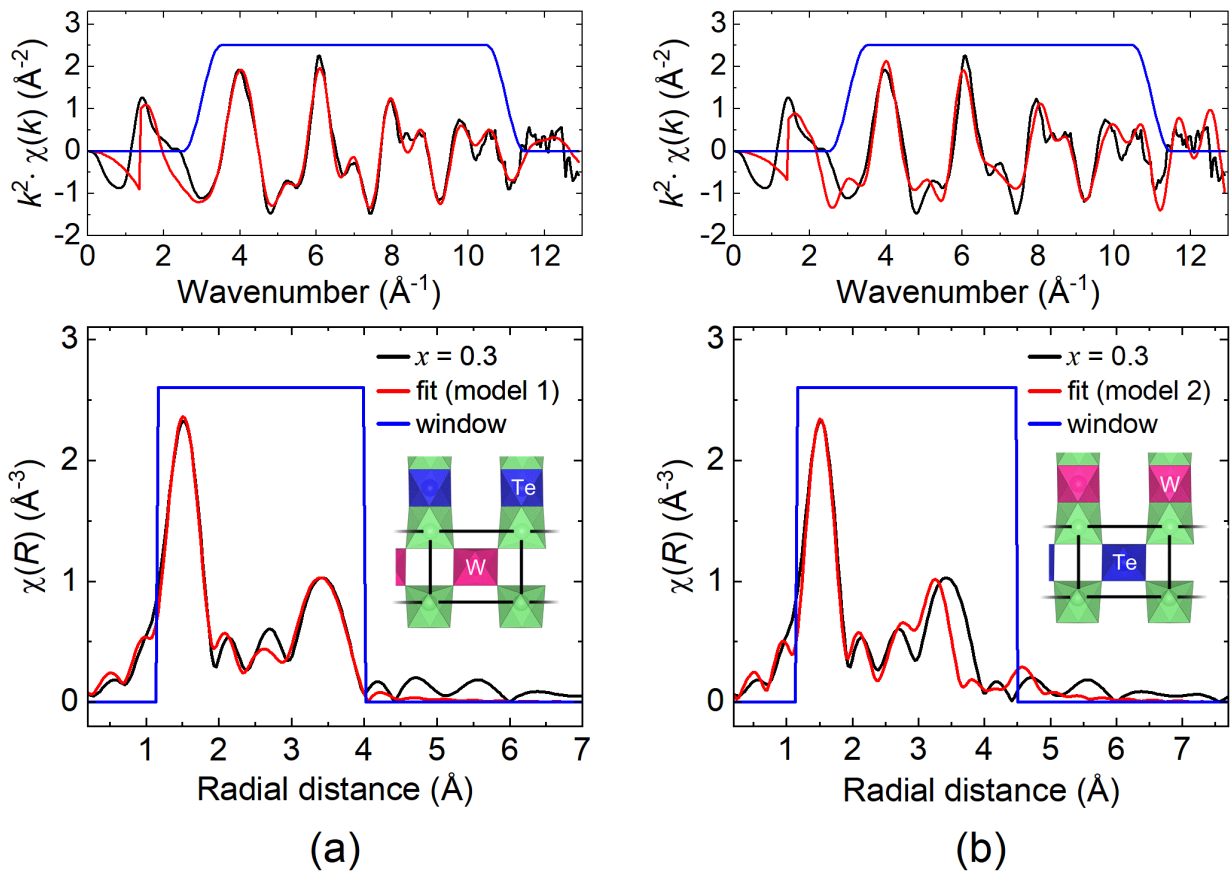

(a)  (b)

**Figure 3.** (a) $k^2\chi(k)$ and $\chi(R)$ W $L_3$ EXAFS data of Ba$_2$CuTe$_{0.7}$W$_{0.3}$O$_6$ with model (1), assuming $W^{6+}$ doping on $B''$(c) site (uncorrected for phase shift). (b) $k^2\chi(k)$ and $\chi(R)$ W $L_3$ EXAFS data of Ba$_2$CuTe$_{0.7}$W$_{0.3}$O$_6$ with model (2), assuming $W^{6+}$ doping on $B''$(f) site (uncorrected for phase shift). In both (a) and (b), the solid black lines represent the experimental data and the red lines represent the model fits. Fitting windows are indicated by solid blue lines. The crystal structures in the $\chi(R)$ vs radial distance plots

depict the models used in the fits. Placement of $W^{6+}$ on the $B''$(c) site in model (1) and $B''$(f) site in model (2) is shown in pink, while $Te^{6+}$ are shown in blue. The $Cu^{2+}$ cations in the spin ladder are green.

**3.1.3 Neutron diffraction** In low dimensional systems, the most striking quantum magnetic behavior may emerge at low temperatures.[47,48] Consequently, variable temperature neutron diffraction studies were performed on $x$ = 0.1 and $x$ = 0.3 between 2-300 K to identify the low temperature structure across the series. Both $x$ = 0.1 and $x$ = 0.3 undergo the same $C2/m$ to $P\bar{1}$ transition as $Ba_2CuTeO_6$ on cooling.[41] The transition ($T_{trans}$), arising from further axial elongation of the Jahn-Teller active Cu-O bonds, is weak meaning the structural integrity of the spin ladder geometry is retained and there is little impact on the magnetic interactions.[38] The transition was marked by peak splitting, as can be observed by comparing the neutron diffraction patterns of $Ba_2CuTe_{0.7}W_{0.3}O_6$ at (a) 1.44 K and (b) 300 K in Figure 4a. As $x$ in $Ba_2CuTe_{1-x}W_xO_6$ increased, peak splitting was suppressed to lower temperatures. Where $T_{trans}$ is just below room temperature (287 K) for $Ba_2CuTeO_6$[41], comparing the $R$-factors from refinements using the $C2/m$ and $P\bar{1}$ models places $T_{trans}$ between 235-240 K in $x$ = 0.1 and 120-140 K in $x$ = 0.3. The decrease in $T_{trans}$ is expected to follow across the series down to the minimum at $x$ = 0.3 as cation disorder increases with $x$. Importantly, it is clear the low temperature structure at 2 K is triclinic.

The unit cell volume ($V_{cell}$) in the $C2/m$ and $P\bar{1}$ phases is plotted as a function of temperature ($T$) in the supplementary Figures S4 and S5. $V_{cell}$ vs $T$ is linear in the $C2/m$ phase, but upon entry to the $P\bar{1}$ phase, the volume deviates from linear behavior for both $x$ = 0.1 and $x$ = 0.3 as shown in Figures S6 and S7. The deviation becomes stronger as the temperature approaches 2 K and is caused by anisotropic thermal expansion of the unit cell. Identical to the monoclinic structure in Figure 1c, the triclinic unit cell consists of face-sharing Cu-$B''$(f)-Cu trimers aligned along the $c$-axis. The Cu-$B''$(f)-Cu trimers are connected by corner sharing $TeO_6$ units creating a 12R stacking sequence. To avoid bringing the $B''$(f)$^{6+}$ and $Cu^{2+}$ cations unfavorably close, the structure avoids reducing the unit cell length along $c$, leading to very little change in the $B''$(f)-Cu distance. Instead, the reduction in the unit cell lengths is directed along $a$ and $b$ by decreasing the corner sharing $B''$(c)-O-Cu angles; the result of which is elongation of the unit cell along $c$. The net effect is a reduction in $V_{cell}$, but to a lesser degree than expected.

The temperature dependence on the $CuO_6$ octahedra was measured empirically using the J-T distortion parameter ($\sigma_{JT}$) in equation 1[49,50].

$$\sigma_{JT} = \sqrt{\frac{1}{6}\sum_i[(Cu-O)_i - \langle Cu-O \rangle]^2} \quad (1)$$

Here, $(\text{Cu} - \text{O})_i$ represents the Cu-O neutron bond length on the $i$-Cu(1)O$_6$ site and $\langle \text{Cu} - \text{O} \rangle$ is the mean bond length. $\sigma_{JT}$ is plotted as a function of $T$ for $x$ = 0.1 and $x$ = 0.2 in Figure 4b, where the $T_{\text{trans}}$ ranges are marked on the graph. The large, non-zero values of $\sigma_{JT}$ reflect the characteristic J-T distortion with the two axial Cu-O bonds elongated in the $d_{z^2}$ direction. Uncharacteristically, the elongation of the axial bond lengths is uneven to accommodate both corner- and face-sharing with the $B''$O$_6$ octahedra. The equatorial bond lengths are approximately equal for any given temperature. The distortion parameter increases slightly with decreasing $T$ for both $x$ = 0.1 and $x$ = 0.3, reflecting further gradual distortion of CuO$_6$ on cooling from 300 to 100 K, with no large discontinuities. There is no further distortion on cooling below 100 K, showing both structures possess the same distortion limit irrespective of W$^{6+}$ content. Within the monoclinic phase, $x$ = 0.3 has a consistently lower value of $\sigma_{JT}$, implying increased W$^{6+}$ doping slightly reduces J-T distortion. The Cu-O bond distances support this with less elongation of the axial Cu-O bonds for the $x$ = 0.3 composition and a smaller variance in the Cu-O bond length (Figure S14). Given the identical W$^{6+}$ and Te$^{6+}$ cation sizes, it is expected this is caused by differences in covalency with the more ionic $d^0$ cations diluting the more covalent nature of the $d^{10}$ cations as $x$ increases.

The similarity in the neutron scattering lengths of Te and W, and the similarity of the preferred coordination site of these cations, meant it was not possible to determine the $B''$(c) vs $B''$(f) site distribution from the neutron diffraction data. Therefore, the $B''$(c) and $B''$(f) site occupancies determined from the synchrotron X-ray diffraction data were used in the crystal structural models.

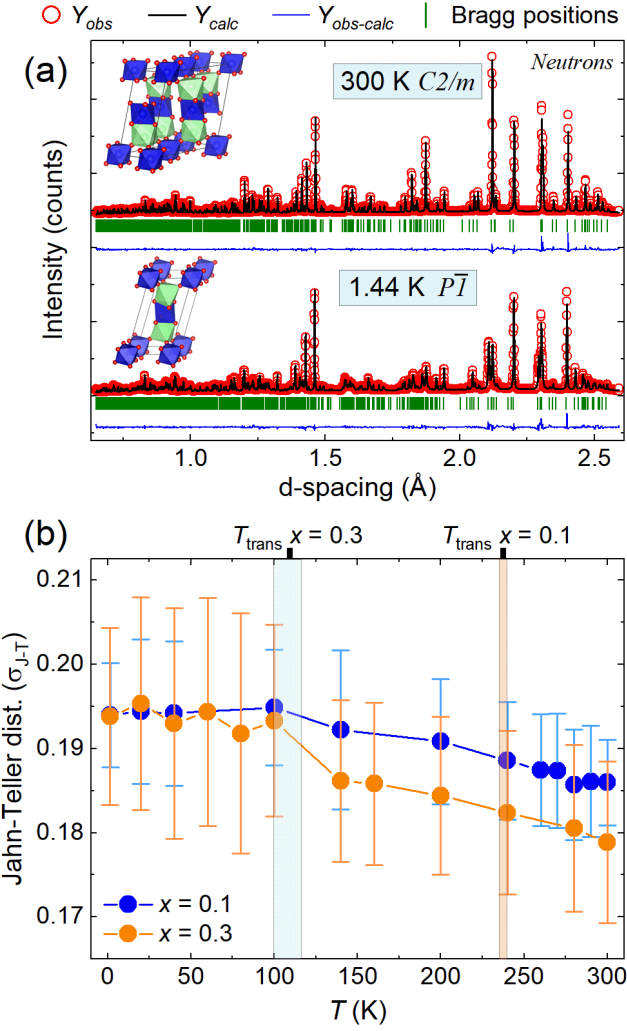

**Figure 4.** Neutron diffraction data showing: (a) High resolution powder diffraction (HRPD) patterns of $Ba_2CuTe_{0.7}W_{0.3}O_6$ at 300 K (top) and 1.44 K (bottom); and (b) The Jahn-Teller distortion parameter ($\sigma_{JT}$) as a function of temperature (*T*) for *x* = 0.1 and *x* = 0.3. The $T_{trans}$ ranges for the monoclinic to triclinic transition on cooling are marked in the plot.

### 3.2 BULK CHARACTERIZATION

**3.2.1 Magnetic susceptibility** DC magnetic susceptibility data collected for *x* = 0, 0.05, 0.1, 0.2 and 0.3 are shown in Figure 5. The $\chi$ vs *T* curve of *x* = 0 is identical to previous reports.[37,39] On cooling there is a broad maximum at $T_{max}$ ~ 74 K, corresponding to short-range ladder interactions. Below $T_{max}$, the susceptibility decreases before leading onto a small upturn beyond 14 K. The low temperature behavior is believed to be indicative of the departure from ladder behavior and entry to the long-range ordered Néel state.[37] However, the low dimensional magnetic behavior means the system does not present a classical AFM ordering cusp. Instead, magnetic order has been detected using muon and neutron techniques placing the magnetic transition at $T_N$ = 14 K.[38,40]

The $W^{6+}$ doped samples are similar in that they all display the same broad $T_{max}$ feature, but the position of $T_{max}$ shifts to lower temperatures as $x$ increases as shown in Table 2, suggesting weakening of the short-range ladder interactions. More dramatic differences are observed at low temperatures, where the upturn in the susceptibility data gradually becomes stronger with $x$ and this 'Curie-tail' like feature is most pronounced in the $x$ = 0.3 sample.

The $\chi$ vs $T$ data between 200-300 K were fitted using the Curie-Weiss law as illustrated in supplementary Figure S19. The Curie constants (C) and Weiss temperatures ($\theta_W$) for each sample are compared in Table 2. Generally, $\theta_W$ becomes more negative as $x$ increases implying stronger AFM interactions. In the tetragonal $(Ba/Sr)_2Cu(Te/W)O_6$ double perovskite systems, the $\theta_W$ is highly correlated with the octahedral rotations, namely the Cu-O-(Te/W) angle.[20] The Cu-O-$B''$(c) angles obtained from the synchrotron X-ray data suggest a similar correlation is present in hexagonal $Ba_2CuTe_{1-x}W_xO_6$, with the bond angles increasing as $\theta_W$ decreases as a function of $x$ (see supplementary Figure S18). The reduction in the Cu-O bond length may also contribute, but the changes between $x$ = 0.1 and $x$ = 0.3 are much less significant compared to the Cu-O-$B''$(c) angle. The value of C was used to calculate the effective magnetic moment $\mu_{eff}$, which in all samples is close to the previously reported value of 1.96 $\mu_B$ per $Cu^{2+}$ for $Cu^{2+}$ in $Ba_2CuTeO_6$.[37] No ZFC and FC divergence was observed in any of the susceptibility curves of the parent and doped samples.

**Table 2.** Results from analysis of DC magnetic susceptibility curves

| $x$ | 0 | 0.05 | 0.1 | 0.2 | 0.3 |
|---|---|---|---|---|---|
| $T_{max}$ (K) | 73.9 | 72.3 | 70.5 | 67.1 | 63.8 |
| C (cm³ K mol⁻¹) | 0.4778(9) | 0.4465(7) | 0.4553(7) | 0.5204(5) | 0.6369(7) |
| $\theta_W$ (K) | -95.1(5) | -86.8(4) | -93.6(4) | -117.6(3) | -151.9(3) |
| $\mu_{eff}$ ($\mu_B$ per $Cu^{2+}$) | 1.95(6) | 1.89(5) | 1.91(5) | 2.04(4) | 2.26(6) |

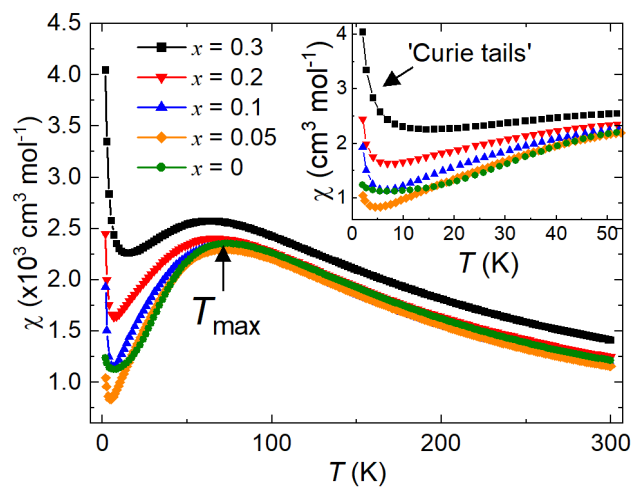

**Figure 5.** DC magnetic susceptibility data for $Ba_2CuTe_{1-x}W_xO_6$ $x = 0 – 0.3$. ZFC curves are shown as a function of temperature $T$ between 2-300 K. The inset shows an expansion of the χ vs $T$ data clearly showing a 'Curie tail' feature when $x > 0$.

**3.2.2 Heat capacity** Zero-field heat capacity ($C_P$) measurements were performed on all samples. The plot in Figure 6a shows the $C_P/T$ data as a function of $T$. Close examination of the curves shows no evidence of an ordering transition in either the parent or doped samples. The former observation agrees with previous heat capacity measurements on $Ba_2CuTeO_6$.[37,39] Owing to weak Néel ordering, $Ba_2CuTeO_6$ possesses strong quantum fluctuations that serve to spread out the magnetic entropy. Hence, $C_P/T$ measurements are largely insensitive to any trace of a lambda ordering peak about $T_N$. The lack of a lambda peak in the $x > 0$ curves show magnetism in the doped samples is similarly weak, as expected from the small $S = 1/2$ $Cu^{2+}$ moment and low dimensional behavior. Hence, the curves appear to be much the same, so it is not possible to distinguish differences in magnetic ordering. However, there are notable trends in the low temperature $C_P/T$ data. The expansion of the range 2-10 K in Figure 6b shows a linear relationship between $C_P/T$ and $T^2$ that is readily fitted using the Debye-Einstein model ($C_P = \gamma T + \beta_D T^3$). The electronic contribution ($\gamma$) to the heat capacity was extracted from the intercept of $C_P/T$ vs $T^2$ and plotted as a function of $x$ in Figure 6c. For $x = 0$, the value of $\gamma$ is almost zero, in excellent agreement with previous studies.[37] However, as $x$ increases, so does the electronic contribution to $C_P$, until the value of $\gamma$ reaches 29.6 mJ mol$^{-1}$ K$^{-2}$ for the $x = 0.3$ sample. Given these materials are Mott insulators, this electronic contribution can only be associated with spins, not conduction electrons.

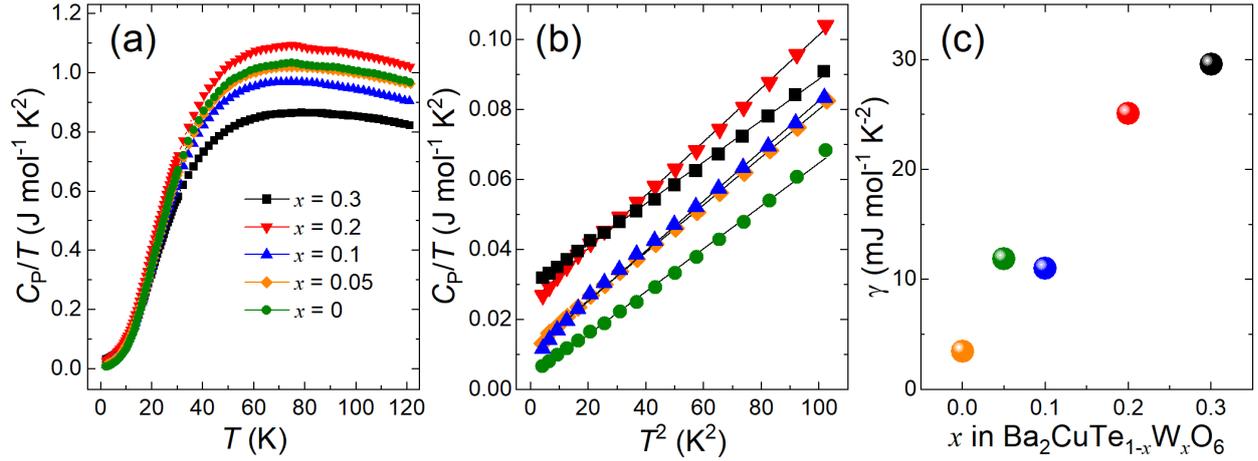

**Figure 6.** Heat capacity data for $Ba_2CuTe_{1-x}W_xO_6$ $x = 0 - 0.3$ including: (a) $C_P/T$ vs $T$ curves for all samples; (b) Debye-Einstein fits of $C_P/T$ vs $T^2$ data between 2-10 K; and (c) Electronic ($\gamma$)-term contribution to $C_P$ as a function of $x$ in $Ba_2CuTe_{1-x}W_xO_6$.

**4. DISCUSSION** The in-depth structural characterization of $Ba_2CuTe_{1-x}W_xO_6$ presented here provides several insights into the structural effects of doping. Variable temperature neutron diffraction shows the disorder introduced by doping affects the $C2/m$ to $P\bar{1}$ transition temperature, with peak splitting occurring at significantly suppressed temperatures as the amount of $W^{6+}$ (and therefore the amount of disorder) increases. The neutron data also show $W^{6+}$ reduces the magnitude of the distortion of the $CuO_6$ octahedra, particularly in the $x = 0.3$ structure. However, the overall changes in bond lengths and angles are minor, which is not surprising given the almost identical $Te^{6+}$ (0.56 Å) and $W^{6+}$ (0.6 Å) ionic radii.[51] Allied with the neutron data, the synchrotron X-ray data go further to reveal a strong preference for the $W^{6+}$ dopant to occupy the $B''(c)$ site in the $Ba_2CuTeO_6$ structure. Refinements show the $W^{6+}$ dopant almost exclusively occupies the corner-sharing $B''(c)$ site. Across the $x = 0.1$, 0.2 and 0.3 compositions, the best fits are obtained when only ~5% of the $W^{6+}$ dopant present in the sample resides on the $B''(f)$ site. This reflects a large difference in favorability between the corner-sharing and face-sharing sites. The EXAFS data corroborates with the observed site preference, providing the best match to the data when the model places $W^{6+}$ exclusively on the $B''(c)$ site.

The strong site preference appears surprising due to the said near-identical ionic radii and identical $W^{6+}$ and $Te^{6+}$ charge.[52] However, it can be explained by considering differences in $Te^{6+}$ and $W^{6+}$ (as well as $Mo^{6+}$) perovskite structures. Perovskites containing $W^{6+}$ and $Mo^{6+}$ exclusively form double perovskite structures, whereas $Te^{6+}$ containing structures can also adopt hexagonal structures.[53] The reason for this lies in differences in metal-oxygen bonding. Owing to the filled $d$-orbitals, metal-oxygen bonding with $Te^{6+}$ has a significant $s$ and $p$ orbital contribution that directs the electron density towards the oxide anions and away from the octahedral face.[53] This reduces the cation-cation repulsion associated with face-sharing octahedra, and such affects will help reduced the energy associated with the Te$^{6+}$ occupation of the *B''*(f) site. The opposite is true for W$^{6+}$ in perovskites where π-bonding interactions lead to highly regular [WO$_6$]$^{6-}$ octahedral units creating a more spherical charge distribution that leads to relatively high repulsion across shared octahedral faces.[54] Consequently, the W$^{6+}$ dopant strongly prefers the corner-sharing *B''*(c) site, where the distance to the Cu$^{2+}$ cation is significantly larger compared to the face sharing *B''*(f) site in the Cu-*B''*(f)-Cu trimer e.g. Te$^{6+}$-Cu$^{2+}$ = 3.962(2) Å (corner-sharing) vs 2.738(1) Å (face-sharing) in *x* = 0.3 at 300 K.

This electrostatic energetic penalty associated with W$^{6+}$ occupancy of face-sharing octahedra is assumed to be responsible for the limited W$^{6+}$ site occupancy on the *B''*(f) site, and reasons why attempts to synthesize richer W$^{6+}$ compositions beyond *x* = 0.3 leads to the formation of increasing quantities of W$^{6+}$ impurities. Furthermore, it explains why Ba$_2$CuTeO$_6$ and Ba$_2$CuWO$_6$ adopt different structures. The Goldschmidt tolerance factor ($t = r_A + r_O/\sqrt{2}(r_B + r_O)$) is used to predict perovskite structures from the ionic radii of the *A* ($r_A$), *B/B'* ($r_B$) and O$^{2-}$ ($r_O$) cations. Given the similar $r_B$, the values of *t* are almost identical for Ba$_2$CuTeO$_6$ (*t* = 1.04) and Ba$_2$CuWO$_6$ (*t* = 1.03), predicting hexagonal or tetragonal symmetry.[53] It is the difference in metal-oxygen bonding that drives W$^{6+}$ to form tetragonal Ba$_2$CuWO$_6$ to maximize the cation distances, while Te$^{6+}$ can accommodate face-sharing in hexagonal Ba$_2$CuTeO$_6$. This highlights the importance of considering more than ionic radii to predict crystal structures. It is worth noting that while seemingly stronger covalency in Te-O may imply stronger superexchange interactions, this is not the case as the *sp* orbitals contribute little to superexchange in these oxides. However, the W$^{6+}$ 5$d^0$ orbitals do participate in the magnetic interactions.[10]

As illustrated in Figure 7, this site preference means W$^{6+}$ doping will mainly tune intra-ladder magnetic interactions in the spin ladder as opposed to between the ladders. Minor occupation of the *B''*(f) site may affect the inter-ladder interactions, but the small occupancy (<5%) is unlikely to have a dramatic effect on the magnetic behavior. Hence, it can be assumed the $J_{leg}$ and $J_{rung}$ interactions directly experience the $d^{10}/d^0$ effect, while the inter-ladder exchange remains unchanged. Predictions of the effect on $J_{leg}$ and $J_{rung}$ can be made based on observations in (Sr/Ba)$_2$Cu(Te/W)O$_6$ which possesses the same structural motif of four corner Cu$^{2+}$ cations interacting via *B'*-O-*B''*-O-*B'* superexchange. The magnetic interactions in Sr$_2$CuTeO$_6$ and tetragonal Ba$_2$CuTeO$_6$ high pressure phase are strongly suppressed as hybridization with W$^{6+}$ 5$d^0$ introduces competing ferromagnetic and antiferromagnetic coupling, leading to a reduced antiferromagnetic $J_2$ exchange in the W$^{6+}$ analogues.[10] Hence, doping W$^{6+}$ on the corner-site is expected to strongly suppress the intra-ladder interactions. There are signs of suppressed ordering, mainly from the large electronic γ-term in the heat capacity data. However, it appears formation of the Néel ordered state is suppressed to a lesser degree, with the γ-term for *x* =

0.3 approaching 50% of the value for $Sr_2CuTe_{0.5}W_{0.5}O_6$.[25] In line with the $d^{10}$ vs $d^0$ effect observed in double perovskites (i.e. $W^{6+}$-strong $J_2$; $Te^{6+}$-strong $J_1$), $W^{6+}$ is also expected to introduce a diagonal component to the intra-ladder interactions, effectively creating twists in the ladder where the $W^{6+}$ are located.

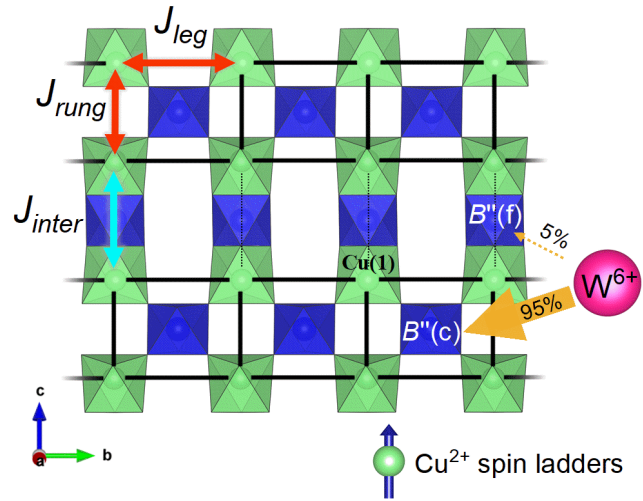

**Figure 7.** Diagram illustrating the strong $W^{6+}$ preference for the corner-sharing $B''(c)$ site versus the face-sharing $B''(f)$ site in the $Ba_2CuTe_{1-x}W_xO_6$ structure. The solid black lines represent the intra-ladder interactions $J_{leg}$ and $J_{rung}$ (red arrows) of the $Cu^{2+}$ spin ladder structure. The dotted lines represent the main $J_{inter}$ inter-ladder interaction (blue arrow) between the spin-ladders. The strong $B''(c)$ site preference means the intra-ladder interactions are most affected by the $W^{6+}$ $d^0$ orbitals.

The results suggest such suppression of the intra-ladder interactions may create a quantum-disordered ground state, but the exact nature of this ground state remains unclear. It is noted the magnetic data possess two main features: (1) $T$-linear low temperature heat capacity and (2) a Curie-tail in the magnetic susceptibility. These features are most pronounced for the higher doped samples, particularly $x$ = 0.3. Although there are number of plausible explanations for the Curie-tail (e.g. orphan spins[55]), the linear low temperature specific heat is a characteristic strongly associated with the random singlet state (RSS).[56] Simply viewed, the RSS state consists of a random lattice covering of singlet dimers (one up spin plus one down spin) and forms due to underlying quenched disorder. Recent studies have suggested quantum-spin liquid behavior in $Sr_2CuTe_{1-x}W_xO_6$ originates from a RSS state induced by square-lattice $W^{6+}$-$Te^{6+}$ disorder.[57–59] Given the structural similarities and comparably large γ-terms, it is possible the observed magnetic behavior for $Ba_2CuTe_{1-x}W_xO_6$ could also originate from a RSS state induced by a random distribution of $Te^{6+}$ $d^{10}$ vs $W^{6+}$ $d^0$ interactions across the $B''(c)$ site. However, experimentally proving the existence of the RSS state is a considerable challenge and would require theoretical studies of the RSS state on a spin ladder, which are yet to be seen.

Before pursuing the RSS state, it is important to first rule out more conventional ground states. A significant γ-term is also consistent with a spin glass.[60] The lack of any frequency dependent features in AC susceptibility measurements would eliminate spin-glass behavior. An alternative possibility is magnetic order. No magnetic Bragg peaks were detected in the HRPD neutron diffraction data for $x$ = 0.1 and $x$ = 0.3, but weak magnetic scattering means formation of a magnetically ordered state cannot be ruled out. Yet, it is unlikely that a magnetically ordered insulator would have such a high electronic contribution to specific heat. Muons would be particularly well-suited to disprove magnetic ordering given their ability to probe local static magnetic fields within the sample.[61] They would also provide further evidence to rule out spin-glass formation from the lack of a spin-glass freezing transition. Another alternative could be formation of a spin-singlet state formed by direct tuning of the intra-ladder interactions. $Ba_2CuTeO_6$ is known to lie on the ordered side of the $S$ = 1/2 spin ladder phase diagram, but is very close to the quantum critical point between the Néel ordered and spin-singlet states.[38,40,62–65] The introduction of $W^{6+}$ $d^0$ to the $B''$(c) site could modify the ratio between the $J_{leg}$ and $J_{rung}$ interactions in such a way that one dominates over the other to form the spin-singlet ground state.[66] In which case, the singlet ground state could be identified from the observation of singlet-triplet excitations in inelastic scattering experiments.[67]

**5. CONCLUSIONS** Chemical doping of the hexagonal perovskite $Ba_2CuTeO_6$ delivers a $Ba_2CuTe_{1-x}W_xO_6$ solid solution (0 ≤ $x$ ≤ 0.3). Structural differences between the $x$ = 0, 0.05, 0.1, 0.2 and 0.3 samples were investigated using a combination of neutron diffraction, synchrotron X-ray diffraction and EXAFS. This revealed a strong preference for the $W^{6+}$ cations to occupy the corner-sharing $B''$(c) site, that sits within the intra-ladder structure of the spin ladder. Site-specific doping results from differences in molecular bonding that leads $W^{6+}$ to prefer the corner-sharing site, where it is further away from $Cu^{2+}$. This site preference leads to direct tuning of the intra-ladder interactions by the $W^{6+}$ $5d^0$ orbitals. In an analogous manner to $Sr_2CuTe_{1-x}W_xO_6$, such direct tuning is expected to suppress the magnetic ordering leading to the formation of a quantum-disordered state.

Further work is needed to uncover the exact nature of the magnetic ground state(s) this creates within the $Ba_2CuTe_{1-x}W_xO_6$ solid solution. However, this demonstrates that the $d^{10}/d^0$ effect can be extended beyond double perovskite structures to modify the magnetic interactions in hexagonal perovskites. Furthermore, the effect could possibly be extended to any structural type that contains corner-sharing octahedra, such as perovskite-derived 2-dimensional structures (e.g. Ruddlesden-Popper phases, Dion-Jacobson phases). This could even be done in a site-selective manner as we have shown here for $Ba_2CuTe_{1-x}W_xO_6$. Our work highlights the $d^{10}/d^0$ effect as a powerful tool for tuning magnetic interactions and ground states in perovskite-derived oxides. This could lead to the discovery of many more materials with exotic magnetic ground states.

**Supporting Information**

The Supporting Information is available free of charge.

Laboratory X-ray diffraction pattern, Rietveld results from neutron diffraction data including bond lengths and angles, Rietveld refinement of synchrotron X-ray diffraction data (file type, Analysis of Extended X-ray Absorption Fine Structure (EXAFS) data, Curie-Weiss fitting of the magnetic susceptibility data. (PDF)

Crystallographic data $Ba_2CuTe_{0.9}W_{0.1}O_6$ 1.55 K (CIF)

Crystallographic data $Ba_2CuTe_{0.9}W_{0.1}O_6$ 300 K (CIF)

Crystallographic data $Ba_2CuTe_{0.7}W_{0.3}O_6$ 1.44 K (CIF)

Crystallographic data $Ba_2CuTe_{0.7}W_{0.3}O_6$ 300 K (CIF)

Crystallographic data $Ba_2CuTe_{0.8}W_{0.2}O_6$ 300 K (CIF)

**Author Contributions**

O.M., C.P., E.J.C. planned and conceived the study, which was supervised by E. J. C.. C.P. synthesized the samples and performed laboratory X-ray diffraction measurements. C.P. and O.M. measured the magnetic susceptibility with help from G.B.G.S.. C.L. and S.E.D. performed heat capacity measurements. O.M., C.P and A.S.G. performed high resolution neutron powder diffraction measurements. M.E. performed synchrotron X-ray diffraction measurements. N.C.H and Bruce Ravel performed the XAS measurements and the EXAFS data obtained was analyzed by C.P. and A.F.. C.P. wrote the manuscript with contributions from all authors.

**Funding Sources**

This work was funded by the Leverhulme Trust Research Project Grant RPG-2017-109.

**ACKNOWLEDGMENTS**

E.J.C., O.M. and C.P. acknowledge financial support from the Leverhulme Trust Research Project Grant No. RPG-2017-109. A.S.G. acknowledges funding through an EPSRC Early Career Fellowship EP/T011130/1. The authors thank the Science and Technology Facilities Council for beamtime allocated at ISIS. The authors are grateful for access to the MPMS3 instrument at the Materials Characterisation Laboratory at ISIS. We acknowledge DESY (Hamburg, Germany), a member of the Helmholtz Association HGF, for the provision of experimental facilities. Parts of this research were carried out at PETRA III beamline P02.1. Components of this research utilized the HADES/MIDAS facility at the University of Sheffield established with financial support from EPSRC and BEIS, under grant EP/T011424/1.[68] Use of the National Synchrotron Light Source II, Brookhaven National Laboratory, was supported by the U.S. Department of Energy, Office of Science, Office of Basic Energy Sciences, under Contract No. DE-AC02-98CH10886 and beamtime proposal number 303200. We are grateful to Bruce Ravel for assistance in acquisition of W $L_3$ XAS data. Heat Capacity measurements were performed using the Advanced Materials Characterisation Suite, funded by EPSRC Strategic Equipment Grant EP/M000524/1. S.E.D. acknowledges funding from the Winton Programme for the Physics of Sustainability (Cambridge) and EPSRC (EP/T028580/1).

**ABBREVIATIONS**

fcc, face-centered cubic; EXAFS, Extended X-ray Absorption Fine Structure, SQUID, superconducting quantum interference device; MPMS, magnetic property measurement system; J-T, Jahn-Teller; PPMS, physical property measurement system; ZFC, zero-field-cooled; FC, field-cooled; XAS, X-ray absorption spectra; RSS, random singlet state.

# Supplementary information

1. **Laboratory X-ray diffraction (Rigaku Miniflex)**

Fig. S1 shows an example laboratory X-ray diffraction pattern of $Ba_2CuTe_{0.9}W_{0.1}O_6$ collected using a Rigaku Miniflex diffractometer (300 K, Cu $K\alpha_1$/ $K\alpha_2$ ($\lambda$ = 1.5405 and 1.5443 Å). The inset shows the unit cell volume change across the solid solution between $x$ = 0 to 0.3.

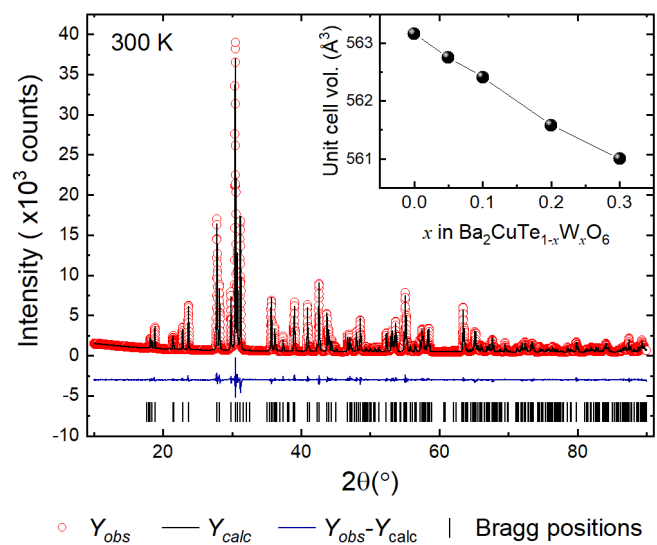

**Fig. S1** Laboratory X-ray diffraction pattern of $Ba_2CuTe_{0.9}W_{0.1}O_6$. The inset shows the unit cell volume change as a function of $x$ in $Ba_2CuTe_{1-x}W_xO_6$.

2. **Neutron diffraction**

Neutron diffraction data were processed using the following method. The raw data from all three banks were normalized against the vanadium standard and experimentally corrected for sample absorption. The three banks were then simultaneously refined using the Rietveld method in GSAS2 with either the monoclinic ($C2/m$) or triclinic ($P\bar{1}$) models.[1] The background was modelled using the Chebyshev function with 10 points. The unit cell volume, atomic positions and UISO values were refined with the USIO values for Te and W on the $B''$(c) and $B''$(f) sites constrained to be equivalent. The refined structural models used the refined $B''$(c) and $B''$(f) site fractions for $x$ = 0.1 and $x$ = 0.3 determined from the synchrotron X-ray diffraction results.

In addition to the above, DIFFA was refined for all banks. Refining additional profile parameters such as DIFFC (lower resolution banks 2 and 3 only) and sigma-1 for all banks, led to a very marginal improvement in the fit. Hence, these parameters were not refined. It was necessary to include sample strain which was modelled using the generalized model. Positive UISO values were obtained at all temperatures throughout the monoclinic phase. However, during the test refinements within the triclinic phase, the Ba(1) site UISO became negative. This is because as the temperature is reduced, the Ba(1) site UISO value becomes very close to zero. Therefore, in the final refinements the Ba(1) and Ba(2) sites UISO values were constrained to be equivalent, producing a stable refinement with positive UISO values for all atoms. Constraining the Ba(1) and Ba(2) sites has negligible effect on the fit quality. The tables below show the refined structures of $Ba_2CuTe_{0.9}W_{0.1}O_6$ and $Ba_2CuTe_{0.7}W_{0.3}O_6$ at ~2 K and 300 K.

Table S1. Refined structure of Ba$_2$CuTe$_{0.9}$W$_{0.1}$O$_6$ at 1.55 K

| Space Group: $P\bar{1}$, No. 2, 1.44 K |||||||
|---|---|---|---|---|---|---|
| $R_P$ = 3.45, $R_{wp}$ = 3.97, $R_{exp}$ = 1.66, $\chi^2$ = 8.94, 111 var. |||||||
| $a$ = 5.7065(2) Å, $b$ = 5.8490(2) Å, $c$ = 10.2555(7) Å |||||||
| $\alpha$ = 108.464(1)°, $\beta$ = 107.0747(8)°, $\gamma$ = 60.7373(3)° |||||||
| Vol. = 279.233(5) Å$^3$ |||||||
| Site | Wyckoff Position | x | y | z | Site fraction | Uiso |
| Ba1 | 2i | 0.13093(40) | 0.12213(35) | 0.38101(19) | 1.0 | 0.00192(32) |
| Ba2 | 2i | 0.27722(40) | 0.28124(34) | 0.85036(18) | 1.0 | 0.0019 |
| Te1 | 1a | 0 | 0 | 0 | 0.809 | 0.00403(53) |
| W1 | 1a | 0 | 0 | 0 | 0.191 | 0.00403 |
| Te2 | 1h | 0.5 | 0.5 | 0.5 | 0.991 | 0.00388(56) |
| W2 | 1h | 0.5 | 0.5 | 0.5 | 0.009 | 0.00388(56) |
| Cu1 | 2i | 0.41345(26) | 0.40522(25) | 0.21466(12) | 1.0 | 0.0055(38) |
| O1 | 2i | 0.62858(35) | 0.58280(40) | 0.37455(17) | 1.0 | 0.00619(47) |
| O2 | 2i | 0.16934(33) | 0.57361(38) | 0.36287(19) | 1.0 | 0.00551(45) |
| O3 | 2i | 0.35992(37) | 0.86776(31) | 0.59887(18) | 1.0 | 0.00793(48) |
| O4 | 2i | 0.79881(42) | 0.78944(41) | 0.90323(18) | 1.0 | 0.00710(44) |
| O5 | 2i | 0.26919(41) | 0.78131(40) | 0.87853(19) | 1.0 | 0.00631(49) |
| O6 | 2i | 0.20133(42) | 0.76452(33) | 0.12686(18) | 1.0 | 0.00790(48) |

Table S2. Refined structure of Ba$_2$CuTe$_{0.9}$W$_{0.1}$O$_6$ at 300 K

| Space Group: $C2/m$, No. 12, 300 K |||||||
|---|---|---|---|---|---|---|
| $R_P$ = 3.10, $R_{wp}$ = 3.85, $R_{exp}$ = 1.22, $\chi^2$ = 19.10, var. 86 |||||||
| $a$ = 10.2278(2) Å, $b$ = 5.72160(4) Å, $c$ = 10.0958(2) Å, $\beta$ = 107.9556(5)° |||||||
| Vol. = 562.03(1) Å$^3$ |||||||
| Site | Wyckoff Position | x | y | z | Site fraction | Uiso |
| Ba1 | 4i | 0.12823(18) | 0 | 0.37994(18) | 1.0 | 0.00553(46) |
| Ba2 | 4i | 0.28256(19) | 0 | 0.84900(18) | 1.0 | 0.01041(50) |
| Te1 | 2a | 0 | 0 | 0 | 0.809 | 0.00704(57) |
| W1 | 2a | 0 | 0 | 0 | 0.191 | 0.00704(57) |
| Te2 | 2d | 0 | 0.5 | 0.5 | 0.991 | 0.00761(56) |
| W2 | 2d | 0 | 0.5 | 0.5 | 0.009 | 0.00761(56) |
| Cu1 | 4i | -0.09421(13) | 0.5 | 0.21445(11) | 1.0 | 0.00981(41) |
| O1 | 4i | 0.13295(17) | 0.5 | 0.40048(18) | 1.0 | 0.01445(50) |
| O2 | 8j | -0.10503(11) | 0.72871(21) | 0.36907(11) | 1.0 | 0.01020(36) |
| O3 | 4i | 0.31753(19) | 0.5 | 0.87535(19) | 1.0 | 0.01884(54) |
| O4 | 8j | 0.04970(14) | 0.76066(27) | 0.88914(13) | 1.0 | 0.01698(40) |

Table S3. Refined structure of $Ba_2CuTe_{0.7}W_{0.3}O_6$ at 1.44 K

| Space Group: $P\bar{1}$, No. 2, 1.44 K |
|---|
| $R_P$ = 4.18, $R_{wp}$ = 4.72, $R_{exp}$ = 1.75, $\chi^2$ = 11.97, 111 var. |
| $a$ = 5.7008(4) Å, $b$ = 5.8402(5) Å, $c$ = 10.223(1) Å |
| $\alpha$ = 108.250(2)°, $\beta$ = 106.693(2)°, $\gamma$ = 60.7631(6)° |
| Vol. = 278.328(9) Å$^3$ |

| Site | Wyckoff Position | x | y | z | Site fraction | Uiso |
|---|---|---|---|---|---|---|
| Ba1 | 2i | 0.13036(72) | 0.12442(51) | 0.38204(29) | 1.0 | 0.00091(45) |
| Ba2 | 2i | 0.28212(74) | 0.28020(49) | 0.85115(27) | 1.0 | 0.00091(45) |
| Te1 | 1a | 0 | 0 | 0 | 0.43 | 0.00363(78) |
| W1 | 1a | 0 | 0 | 0 | 0.57 | 0.00363(78) |
| Te2 | 1h | 0.5 | 0.5 | 0.5 | 0.97 | 0.00354(79) |
| W2 | 1h | 0.5 | 0.5 | 0.5 | 0.03 | 0.00354(79) |
| Cu1 | 2i | 0.40939(45) | 0.4028(38) | 0.21492(18) | 1.0 | 0.00605(54) |
| O1 | 2i | 0.62719(60) | 0.58563(71) | 0.37253(29) | 1.0 | 0.00537(76) |
| O2 | 2i | 0.16580(58) | 0.57569(67) | 0.36446(31) | 1.0 | 0.00490(73) |
| O3 | 2i | 0.36352(71) | 0.86942(46) | 0.59997(28) | 1.0 | 0.01037(70) |
| O4 | 2i | 0.80705(75) | 0.78764(76) | 0.89856(34) | 1.0 | 0.00751(74) |
| O5 | 2i | 0.28312(74) | 0.78123(74) | 0.88446(36) | 1.0 | 0.00682(79) |
| O6 | 2i | 0.19415(73) | 0.76774(51) | 0.12851(27) | 1.0 | 0.00822(73) |

Table S4. Refined structure of $Ba_2CuTe_{0.7}W_{0.3}O_6$ at 300 K

| Space Group: $C2/m$, No. 12, 300 K |
|---|
| $R_P$ = 3.35, $R_{wp}$ = 3.84, $R_{exp}$ = 1.74, $\chi^2$ = 11.02, var. 86 |
| $a$ = 10.2118(3) Å, $b$ = 5.71745(5) Å, $c$ = 10.0866(3) Å, $\beta$ = 107.9193(6)° |
| Vol. = 560.35(1) Å$^3$ |

| Site | Wyckoff Position | x | y | z | Site fraction | Uiso |
|---|---|---|---|---|---|---|
| Ba1 | 4i | 0.12841(23) | 0 | 0.38015(22) | 1.0 | 0.00394(55) |
| Ba2 | 4i | 0.28318(25) | 0 | 0.84944(23) | 1.0 | 0.01013(63) |
| Te1 | 2a | 0 | 0 | 0 | 0.43 | 0.00850(75) |
| W1 | 2a | 0 | 0 | 0 | 0.57 | 0.00850(75) |
| Te2 | 2d | 0 | 0.5 | 0.5 | 0.97 | 0.00865(70) |
| W2 | 2d | 0 | 0.5 | 0.5 | 0.03 | 0.00865(70) |
| Cu1 | 4i | -0.09455(17) | 0.5 | 0.21486(14) | 1.0 | 0.01123(51) |
| O1 | 4i | 0.13336(23) | 0.5 | 0.39967(23) | 1.0 | 0.01726(64) |
| O2 | 8j | -0.10543(14) | 0.72904(27) | 0.36959(14) | 1.0 | 0.01200(45) |
| O3 | 4i | 0.31705(24) | 0.5 | 0.87607(23) | 1.0 | 0.01712(65) |
| O4 | 8j | 0.04946(17) | 0.76007(34) | 0.88959(16) | 1.0 | 0.01577(48) |

a) Monoclinic to triclinic phase transition

The $C2/m$ to $P\bar{1}$ transition temperature ($T_{trans}$) was determined by plotting the global $R_{wp}$ and $\chi^2$ as a function of temperature (Fig. S2 and Fig. S3). The lower $R_{wp}$ and high $\chi^2$ at base (1.55 K or 1.44 K), 100 K, 200 K and 300 K reflects the longer counting time used for these datasets. On cooling, the rise in the value of $R_{wp}$ and $\chi^2$ shows the failure of the monoclinic model. Below this temperature, the triclinic model become a better description of the structure. $T_{trans}$ occurs between 240-235 K in $Ba_2CuTe_{0.9}W_{0.1}O_6$ and between 100-120 K in $Ba_2CuTe_{0.7}W_{0.3}O_6$.

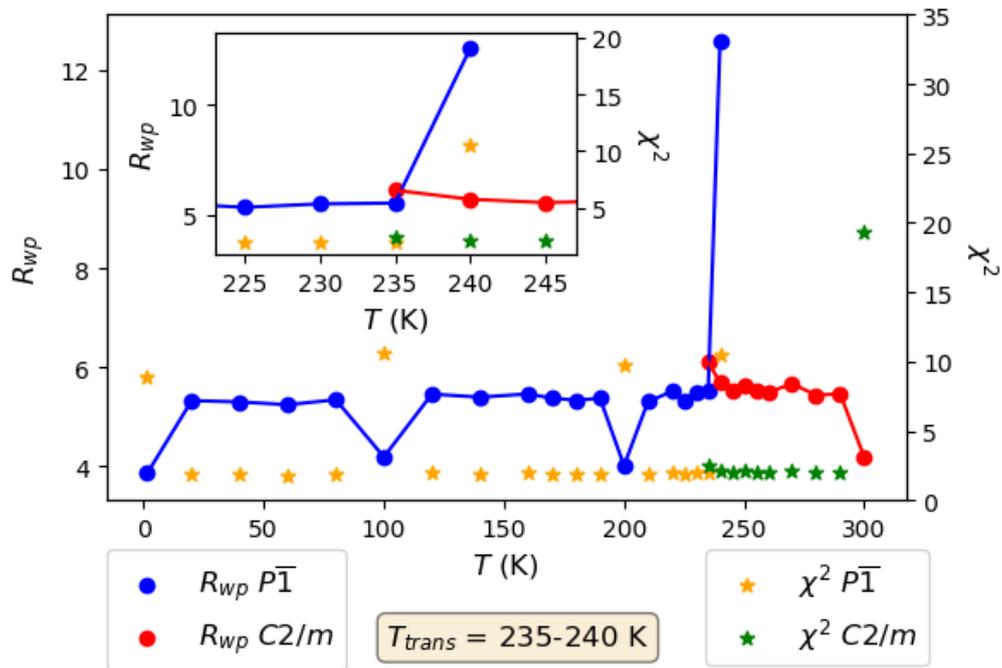

**Fig. S2** *R*-values vs *T* for $Ba_2CuTe_{0.9}W_{0.1}O_6$ neutron data. The left y-axis shows $R_{wp}$ and the right y-axis shows $\chi^2$ as a function of temperature, *T*.

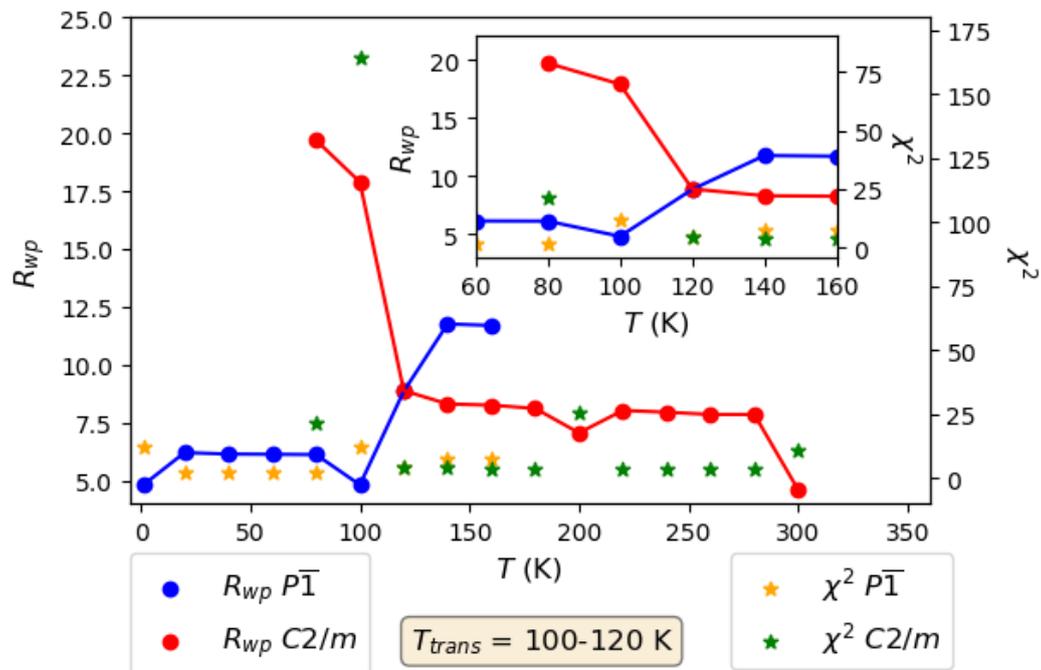

**Fig. S3** *R*-values vs *T* for $Ba_2CuTe_{0.7}W_{0.3}O_6$ neutron data. The left y-axis shows $R_{wp}$ and the right y-axis shows $\chi^2$ as a function of temperature, *T*.

b) Lattice parameters determined from Rietveld analysis of neutron diffraction data.

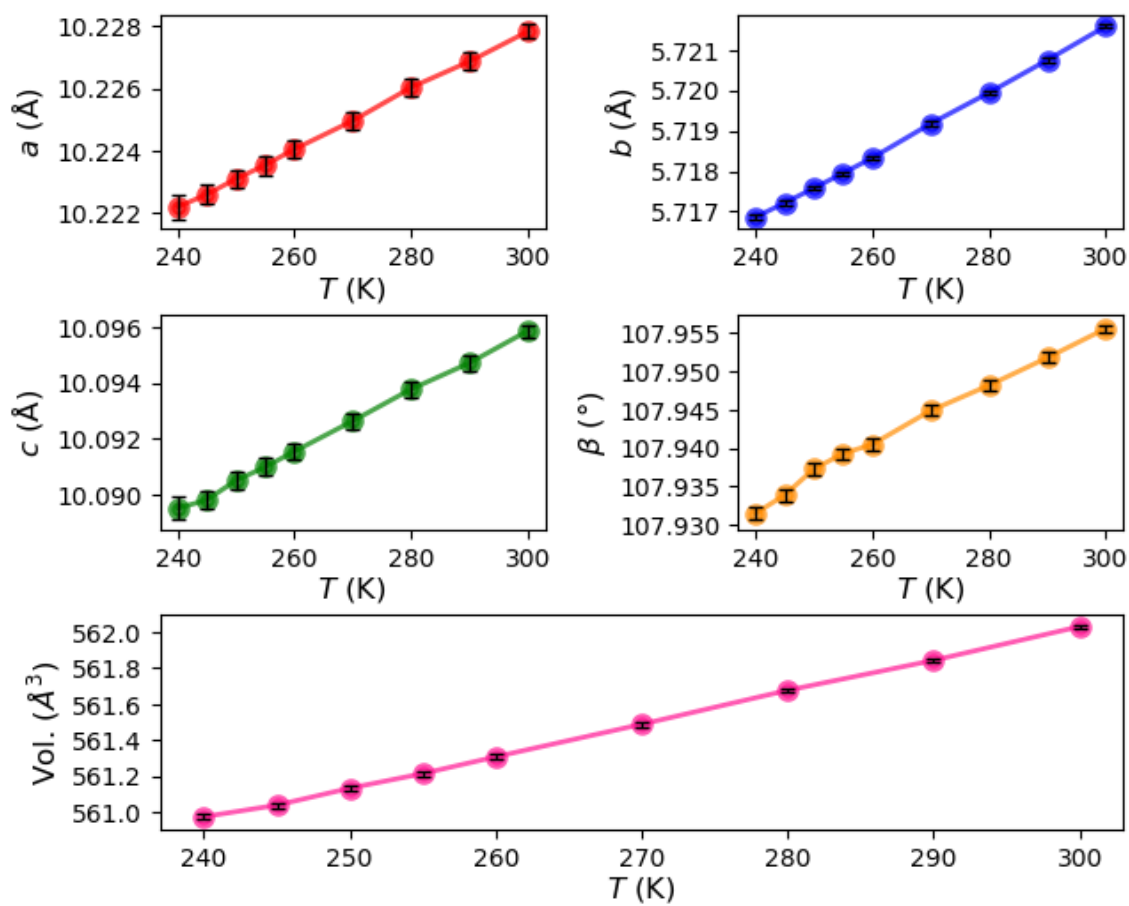

**Fig. S4** Unit cell parameters and unit cell volume of $Ba_2CuTe_{0.9}W_{0.1}O_6$ in the monoclinic phase, $C2/m$, as a function of $T$ between 240-300 K.

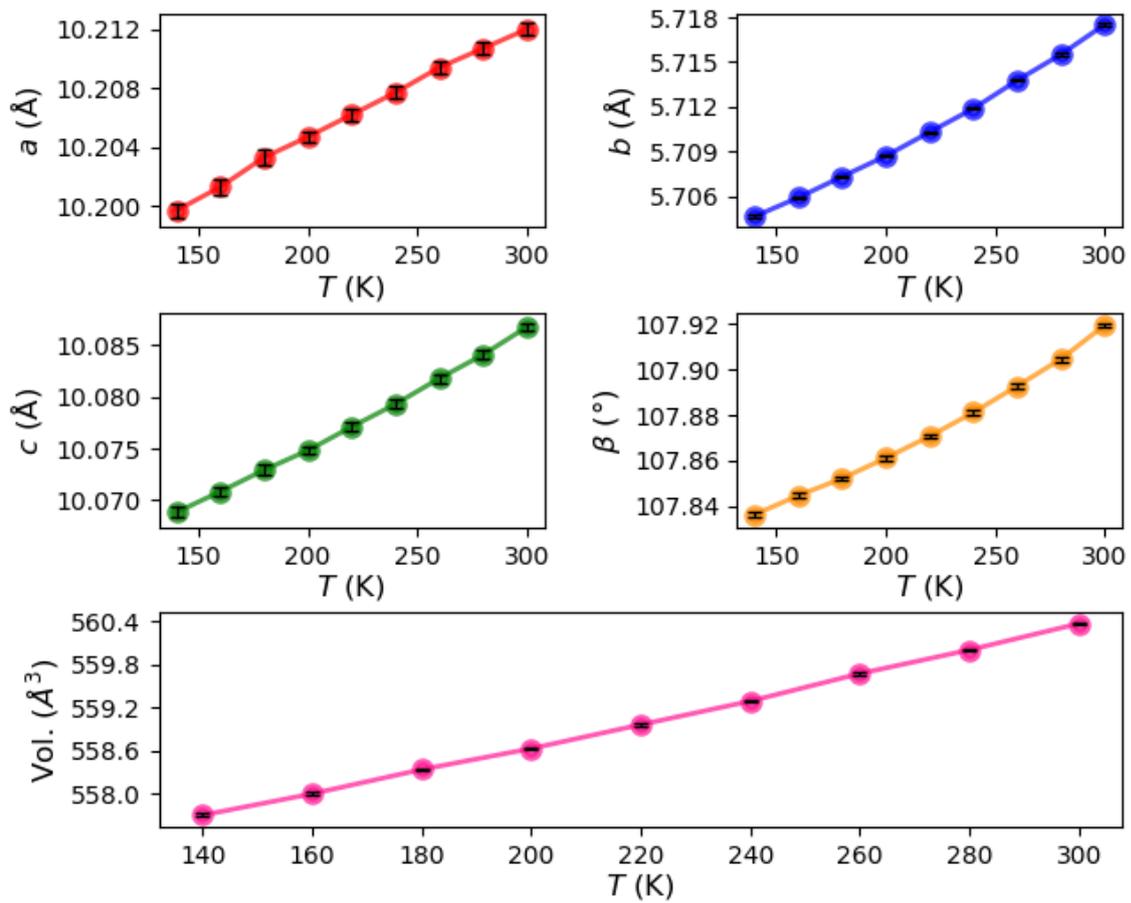

**Fig. S5** Unit cell parameters and unit cell volume of Ba$_2$CuTe$_{0.7}$W$_{0.3}$O$_6$ in the monoclinic phase, $C2/m$, as a function of $T$ between 140-300 K.

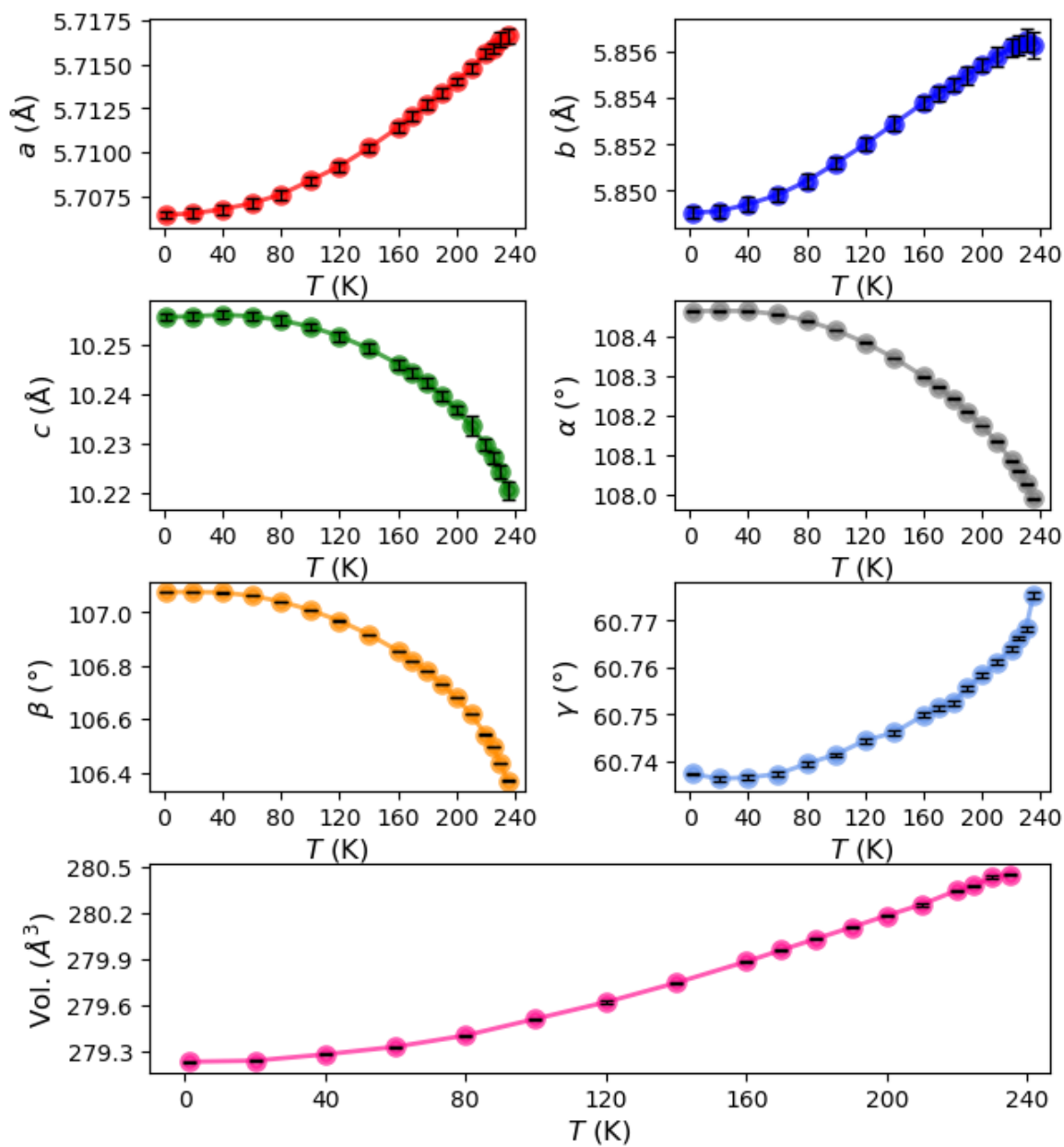

**Fig. S6** Unit cell parameters and unit cell volume of Ba$_2$CuTe$_{0.9}$W$_{0.1}$O$_6$ in the triclinic phase, $P\bar{1}$, as a function of $T$ between 1.55-235 K.

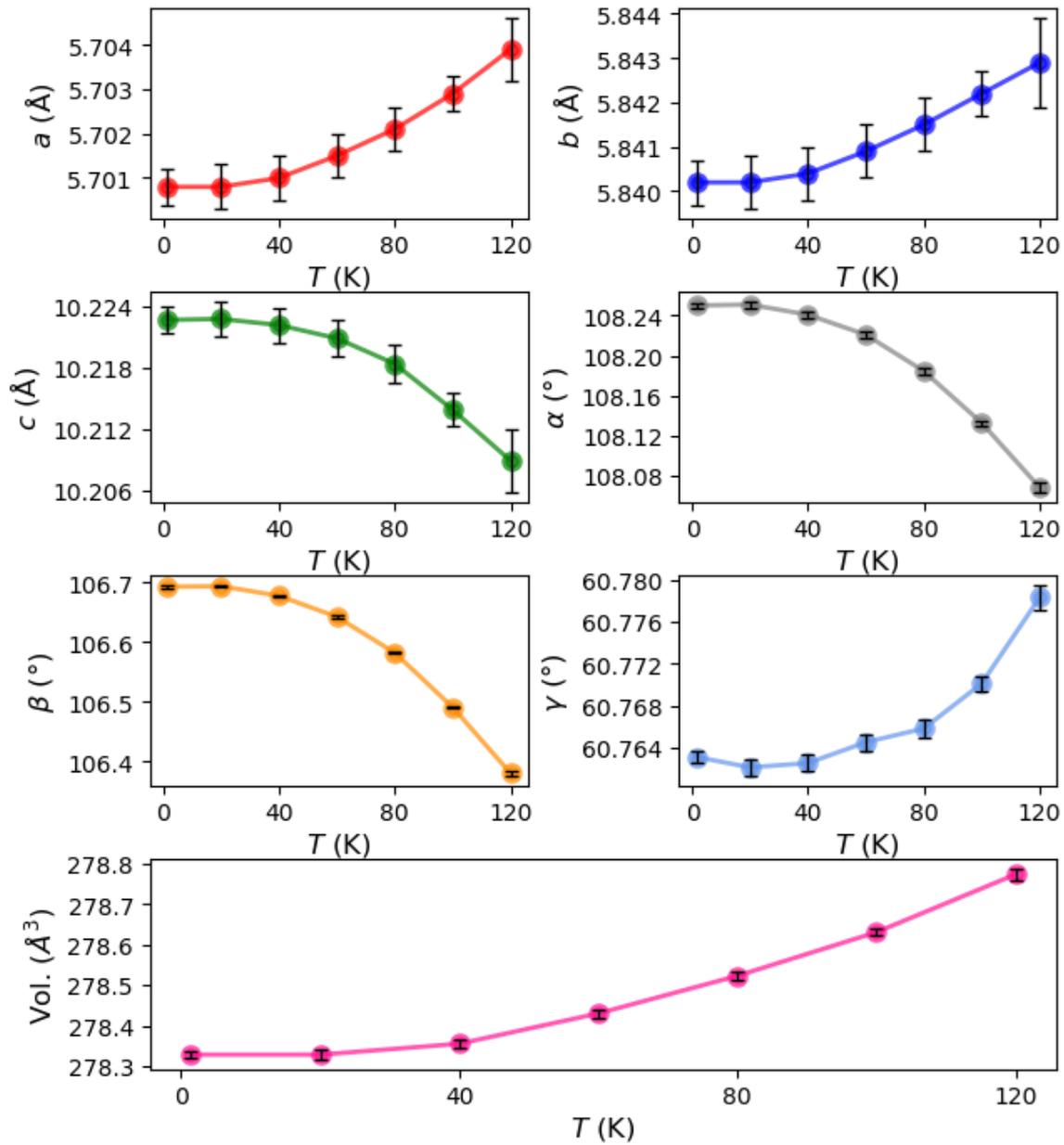

**Fig. S7** Unit cell parameters and unit cell volume of $Ba_2CuTe_{0.7}W_{0.3}O_6$ in the triclinic phase, $P\bar{1}$, as a function of $T$ between 1.44-120 K.

b) Non-linear unit cell volume in triclinic phase

The unit cell volume changes non-linearly in the triclinic phase of both $Ba_2CuTe_{0.9}W_{0.1}O_6$ and $Ba_2CuTe_{0.7}W_{0.3}O_6$. Several possible reasons for this behaviour were investigated. Firstly, Jahn-Teller distortion of the $CuO_6$ octahedra within the triclinic phase was determined with temperature. Fig. S8 and S9 show little change in the distortion with temperature, with the Jahn-Teller distortion parameter ($\sigma_{JT}$) remaining approximately constant with $T$ in both $x$ = 0.1 and 0.3.

Jahn-Teller distortion parameter: $\sigma_{JT} = \sqrt{\frac{1}{6}\Sigma_i[(Cu-O)_i - \langle Cu-O \rangle]^2}$

The Te(1)-Cu(1) and Te(2)-Cu(1) distances were then investigated. Fig. S10 and S11 show the distances for $x$ = 0.1 and $x$ = 0.3 respectively. It was observed in both $x$ = 0.1 and 0.3, the Te(1)-Cu(1) distances decreased with *T* which would contribute to a reduction in *a* and *b*, however not significantly enough to account for the total unit cell contraction in the *ab* plane. Similarly, it was observed there is very little change in the Te(2)-Cu(1) distance with temperature. In fact, the slight decrease in the Te(2)-Cu(1) distances within the Cu(1)-Te(2)-Cu(1) trimer should contributed towards a reduction in *c*.

The revelation comes from looking at the bond angle for the three Te/W(1)-O-Cu(1) corner-sharing linkers and the Cu(1)-O-Te(2) bond angle in the face-sharing Cu-Te-Cu timer. The bond angles for $x$ = 0.1 and 0.3 are shown in Fig. S12 and S13. The Te/W(1)-O-Cu(1) bond angles are already significantly perturbed away from 180° in both the monoclinic and triclinic unit cells. However, as *T* is lowered, the distortion gets worse. The angle data for $Ba_2CuTe_{0.9}W_{0.1}O_6$ in Fig. S12 shows the Cu(1)-O(5)-Te(1) bond angle decreases from 171.5° to 168.0° between 200-1.5 K - a change of Δ(°) ~ 3.4. The Cu(1)-O(4)-Te(1) bond angle is reduced by a similar angle of Δ(°) ~ 3.0 within this temperature range, while the Cu(1)-O(6)-Te(1) is reduced less by Δ(°)~ 0.8.

Alternatively, the Cu(1)-O(1-3)-Te(2) bond angles within the Cu-Te-Cu trimer change little with temperature. Reducing the Cu(1)-O-Te(2) bond angles would bring the positively charged face-sharing $Te^{6+}$ and $Cu^{2+}$ cations closer together. This would create a significant repulsion between the cation sites. Hence, the structure prefers to keep the face-sharing cations apart and minimize the reduction of the Te(2)-Cu(1) trimer distance. To do this, the face-sharing trimer acts like a rigid pole along the *c* axis that resists compression as *T* lowers. Instead, the reduction in the unit cell lengths is directed perpendicular to the trimer along *a* and *b* by reducing the Te/W(1)-O-Cu(1) bond lengths and decreasing Te/W(1)-O-Cu(1) bond angles. The combined effect of both pushes the *ab* planes at the top and bottom of the triclinic unit cell up and down, respectively. Whilst this occurs, *a* and *b* decrease, and the unit cell is elongated along *c*. The Te/W(1)-O-Cu(1) bond angles do not alter by the same amount, with the Te/W(1)-O(4, 5)-Cu(1) angles decreasing more than the Te/W(1)-O(6)-Cu(1) angle. This has the effect of reducing γ but increasing β and α as the temperature is lowered. Therefore, the mixture of face- and corner-sharing leads to anisotropic thermal expansion of the unit cell. The competition between contraction along *a* and *b* and expansion along *c*, leads to a non-linear decrease in the unit cell volume of $x$ = 0.1 and $x$ = 0.3 with *T*.

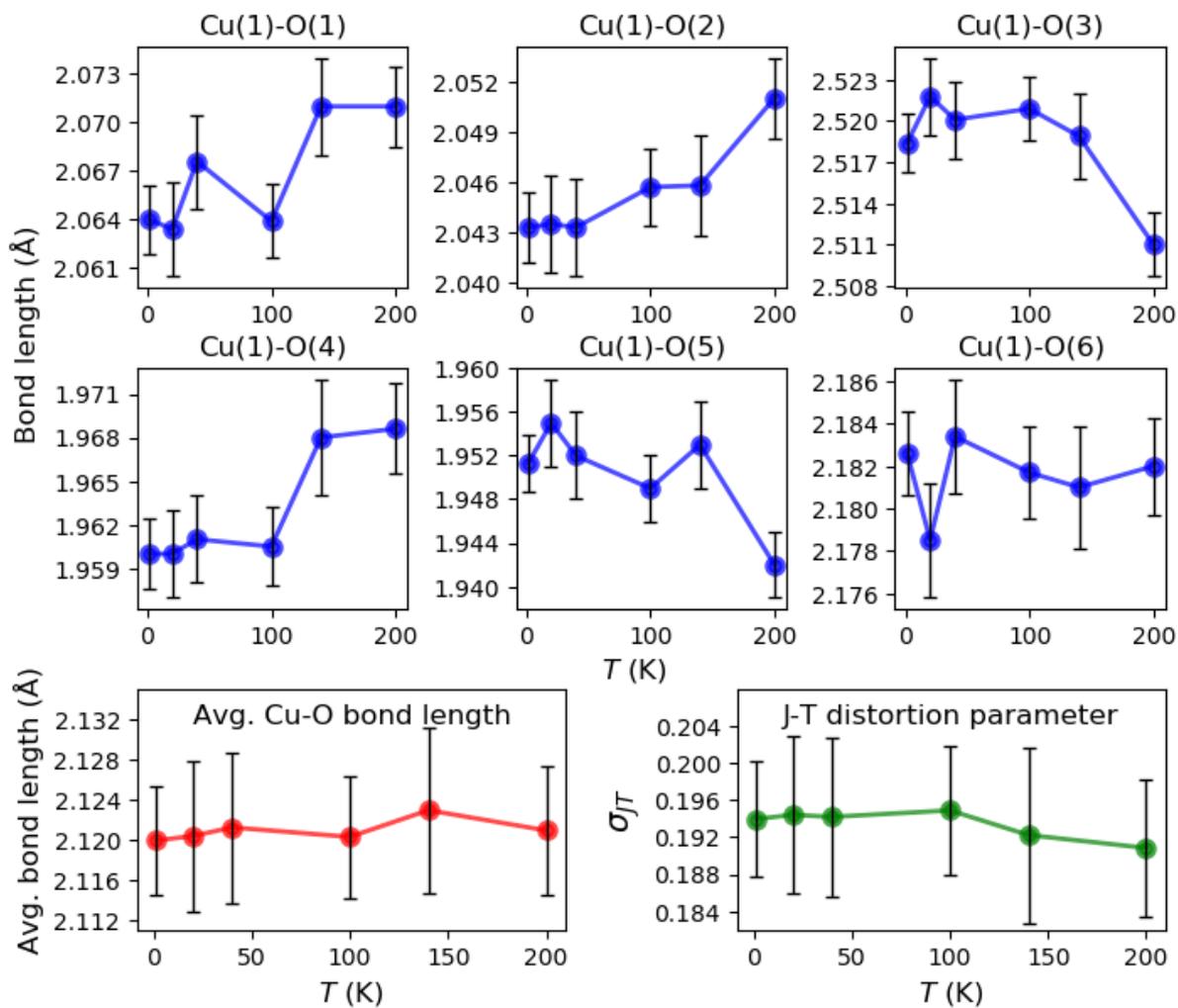

**Fig. S8** Cu(1)-O(1-6) bond lengths in Ba$_2$CuTe$_{0.9}$W$_{0.1}$O$_6$ as a function of $T$ in the $P\bar{1}$ phase. Also shown are the average Cu-O bond length and the Jahn-Teller distortion parameter.

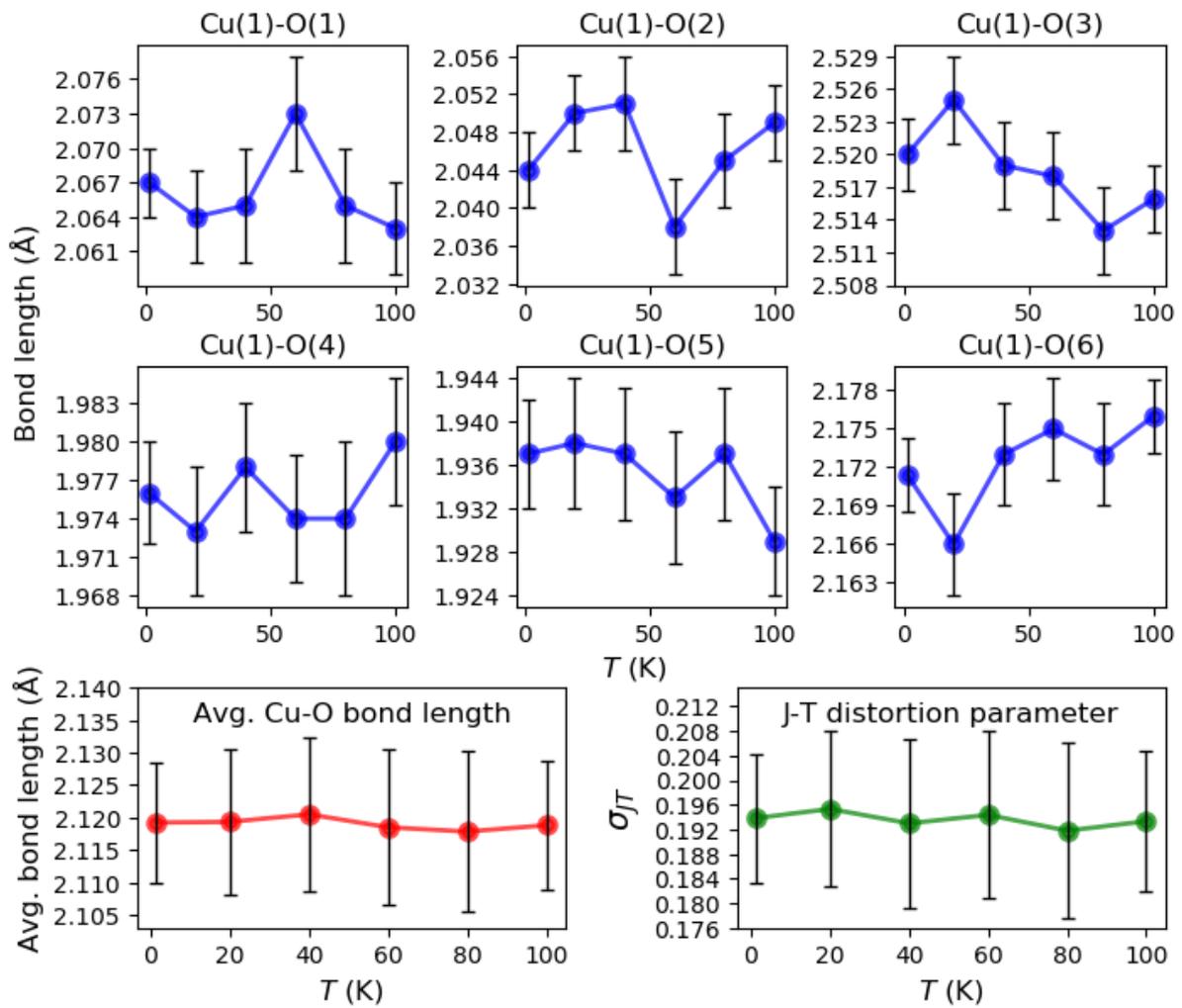

**Fig. S9** Cu(1)-O(1-6) bond lengths in Ba2CuTe$_{0.7}$W$_{0.3}$O$_6$ as a function of $T$ in the $P\bar{1}$ phase. Also shown are the average Cu-O bond length and the Jahn-Teller distortion parameter.

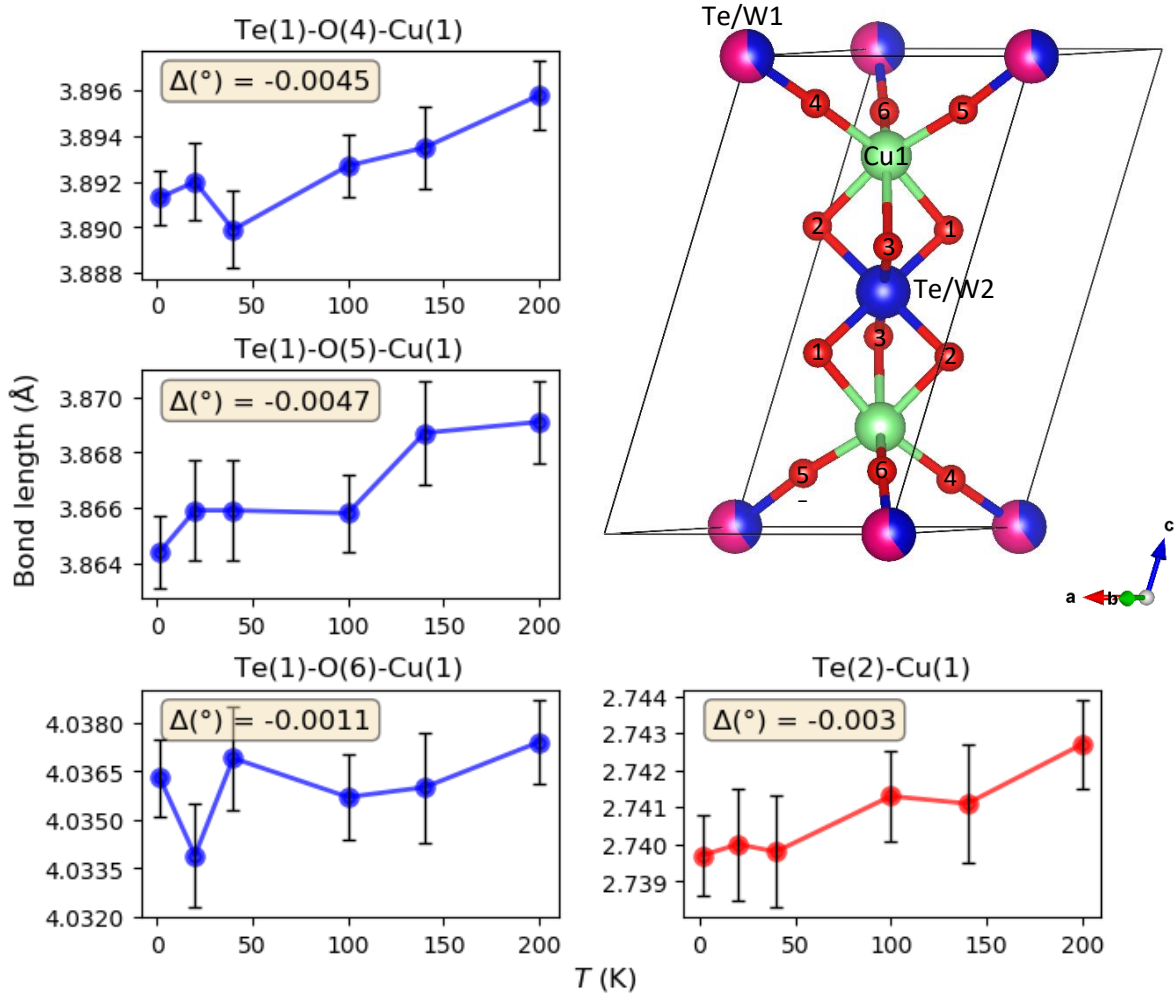

**Fig. S10** Te(1)-Cu(1) and Te(2)-Cu(1) distances in Ba$_2$CuTe$_{0.9}$W$_{0.1}$O$_6$ in the $P\bar{1}$ phase.

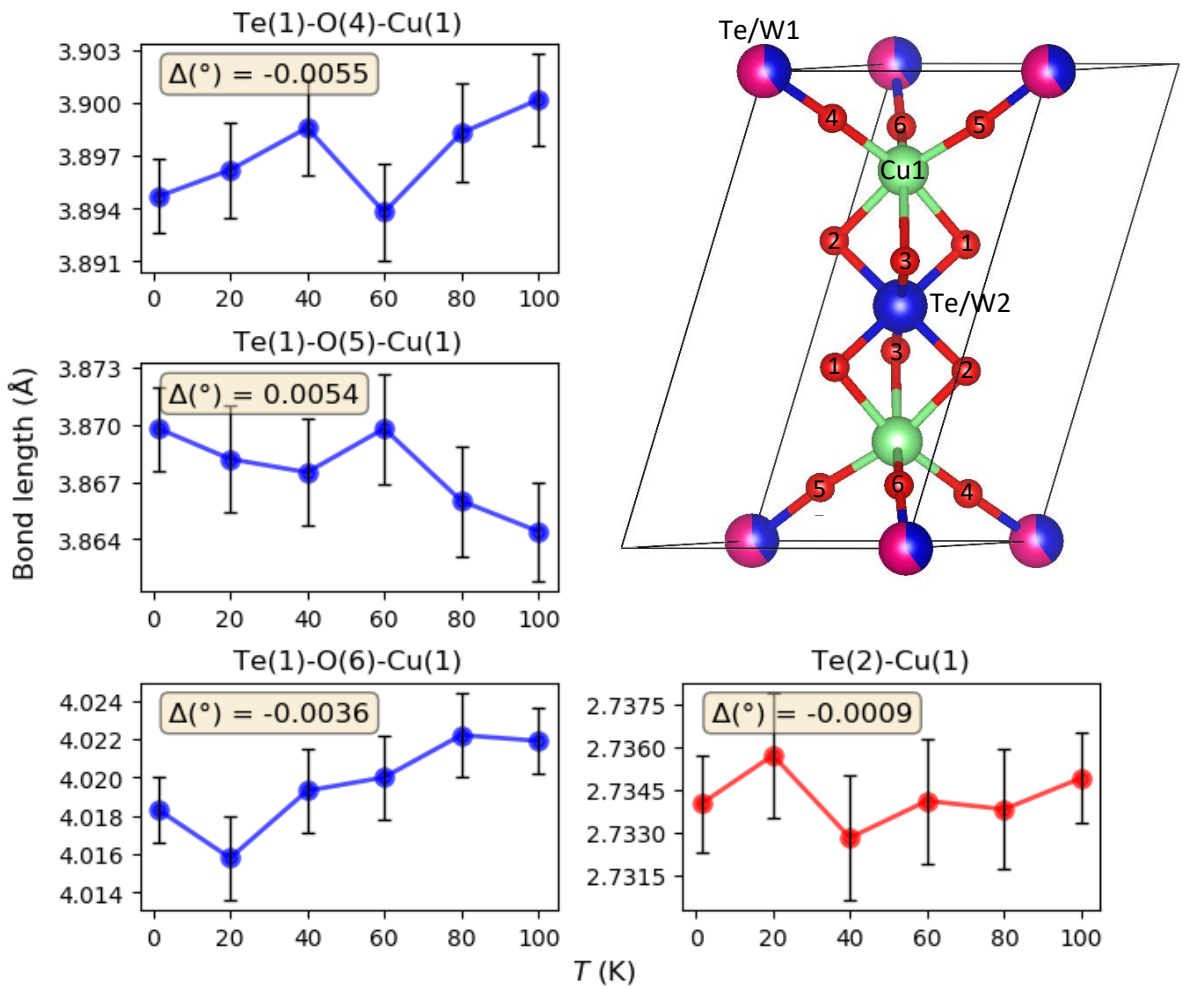

**Fig. S11** Te(1)-Cu(1) and Te(2)- Cu(1) distances in Ba$_2$CuTe$_{0.7}$W$_{0.3}$O$_6$ in the $P\bar{1}$ phase.

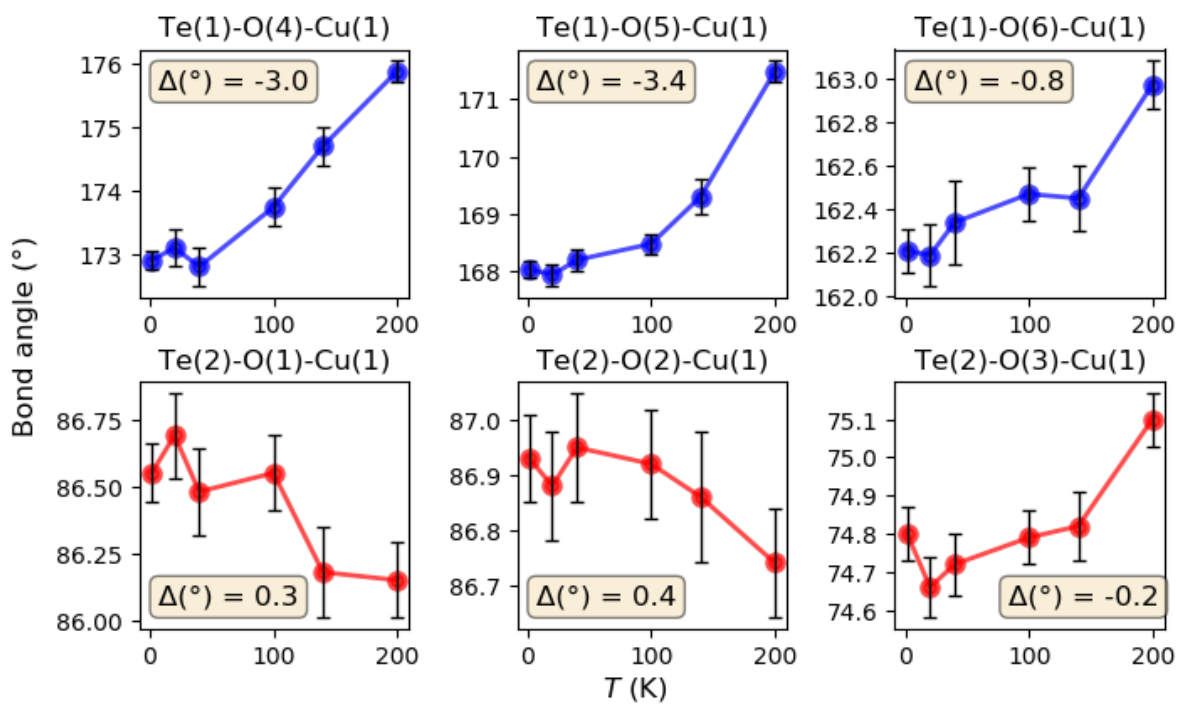

**Fig. S12** Te(1, 2)-O(6-1)-Cu(1) bond angles in Ba$_2$CuTe$_{0.9}$W$_{0.1}$O$_6$ in the $P\bar{1}$ phase.

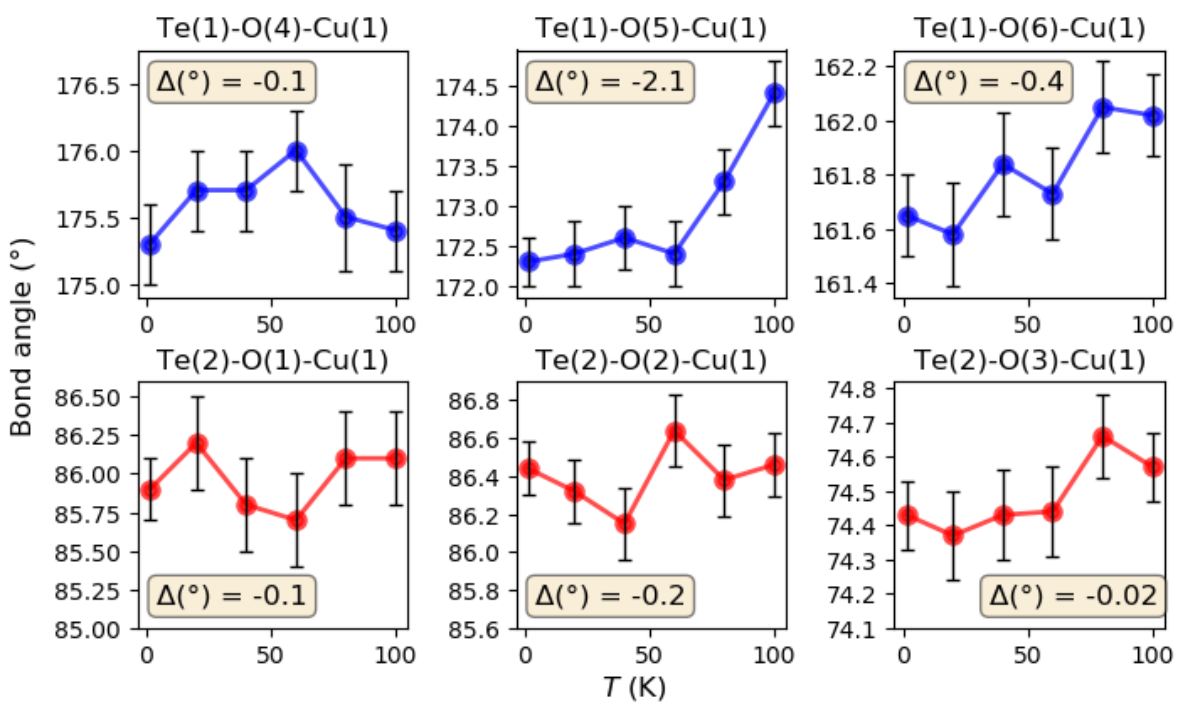

**Fig. S13** Te(1, 2)-O(6-1)-Cu(1) bond angles in Ba$_2$CuTe$_{0.7}$W$_{0.3}$O$_6$ in the $P\bar{1}$ phase.

c) CuO$_6$ octahedral bond lengths in Ba$_2$CuTe$_{0.9}$W$_{0.1}$O$_6$ and Ba$_2$CuTe$_{0.7}$W$_{0.3}$O$_6$

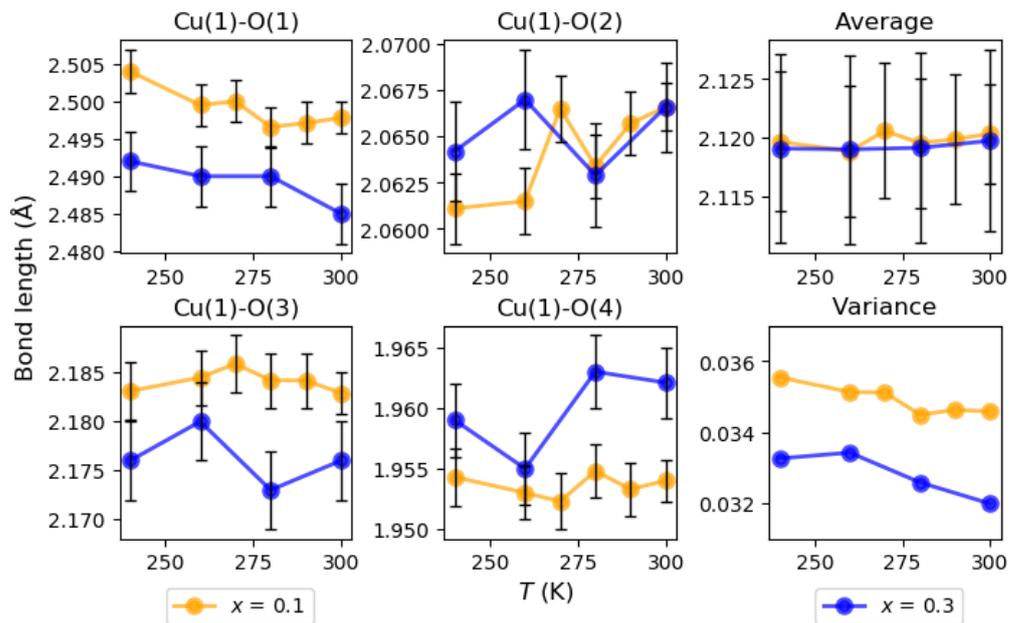

**Fig. S14** Cu(1)-O(1-4) bond lengths in Ba$_2$CuTe$_{0.9}$W$_{0.1}$O$_6$ and Ba$_2$CuTe$_{0.9}$W$_{0.1}$O$_6$ in the $C/2m$ phase.

### 3. Synchrotron X-ray diffraction

Synchrotron X-ray diffraction patterns were collected using a wavelength of $\lambda = 0.20742$ Å at 300K for the following three samples: $Ba_2CuTe_{0.7}W_{0.3}O_6$, $Ba_2CuTe_{0.8}W_{0.2}O_6$ and $Ba_2CuTe_{0.9}W_{0.1}O_6$. To determine the $W^{6+}$ site occupancy, the refinements were performed using three possible monoclinic models. These models were: (1) $W^{6+}$ exclusively on the $B''(c)$ site; (2) $W^{6+}$ exclusively on the $B''(f)$ site; and (3) $W^{6+}$ shared equally between the $B''(c)$ and $B''(f)$ sites. Tables S5, S6 and S7 below show the $R$-values obtained for each sample using these models. For every sample, the model with the lowest $R$-values is model (1) where $W^{6+}$ exclusively occupies the $B''(c)$ corner sharing site. Occupancy of the $B''(f)$ site by $W^{6+}$ produces a poorer fit in comparisons and leads to negative UISO values indicating an unstable refinement.

The $B''(c)$ and $B''(f)$ site fractions in model (1) were refined using the appropriate constraints for the structural composition. The highly correlated UISOs for the $B''(c)$ and $B''(f)$ sites were turned off during the refinement. During the refinement, a small fraction of $W^{6+}$ moved onto the $B''(f)$ site. For each composition, the percentage of the total amount of $W^{6+}$ in $Ba_2CuTe_{1-x}W_xO_6$ residing on the $B''(f)$ site was calculated by dividing the $B''(f)$-W2 site fraction by the $B''(c)$-W1 site fraction. The percentage of the total amount of $W^{6+}$ residing on the $B''(f)$ site was found to be: 4.7(2)% ($Ba_2CuTe_{0.8}W_{0.2}O_6$); 4.7(2)% ($Ba_2CuTe_{0.8}W_{0.2}O_6$); and 5.3(2)% ($Ba_2CuTe_{0.7}W_{0.3}O_6$). Tables S8, S9 and S10 show the refined crystal structures. Placing ~ 5% of the total amount of $W^{6+}$ on the $B''(f)$ site provides a slightly improved description of the crystal structure for $x = 0.2$ and $x = 0.3$, with the $R_{wp}$ values slightly reduced, however the $R$-values for $x = 0.1$ are the same. To check the stability of the refinement, the refined site fractions were changed to those of model (2) i.e. exclusive occupancy of the $B''(f)$ site. Repeating the refinement converged to the same result as when approaching the refinement using site fractions from model (1) in the initial structural model. Overall, this result shows $W^{6+}$ doping strongly favours the $B''(c)$ site, which has a solubility of 95%, while the $B''(f)$ site can only accommodate 5% doping across the solid solution. The inability to achieve phase purity beyond $x = 0.3$ shows the solubility limit for $W^{6+}$ doping of $Ba_2CuTe_{1-x}W_xO_6$ lies between $x = 0.35$ and $x = 0.4$.

Table S5: R-values obtained from refinement of synchrotron X-ray data for $Ba_2CuTe_{0.7}W_{0.3}O_6$ using different $W^{6+}$ site occupancy models

| Site occupancy | $R_{wp}$ | $R_p$ | $R_{exp}$ | $\chi^2$ |
| --- | --- | --- | --- | --- |
| $W^{6+}$ exclusively on $B''(c)$ site | 2.65 | 1.88 | 0.81 | 10.69 |
| $W^{6+}$ exclusively on $B''(f)$ site | 4.81 | 3.35 | 0.81 | 35.28 |
| $W^{6+}$ equal (50:50) on both $B''(c)$ and $B''(f)$ sites | 3.14 | 2.25 | 0.81 | 14.98 |

Table S6: R-values obtained from refinement of synchrotron X-ray data for $Ba_2CuTe_{0.8}W_{0.2}O_6$ using different $W^{6+}$ site occupancy models

| Site occupancy | $R_{wp}$ | $R_p$ | $R_{exp}$ | $\chi^2$ |
|---|---|---|---|---|
| $W^{6+}$ exclusively on $B''$(c) site | 1.76 | 1.34 | 0.82 | 4.58 |
| $W^{6+}$ exclusively on $B''$(f) site | 3.10 | 2.23 | 0.82 | 14.29 |
| $W^{6+}$ equal (50:50) on both $B''$(c) and $B''$(f) sites | 2.07 | 1.57 | 0.82 | 6.35 |

Table S7: R-values obtained from refinement of synchrotron X-ray data for $Ba_2CuTe_{0.9}W_{0.1}O_6$ using different $W^{6+}$ site occupancy models

| Site occupancy | $R_{wp}$ | $R_p$ | $R_{exp}$ | $\chi^2$ |
|---|---|---|---|---|
| $W^{6+}$ exclusively on $B''$(c) site | 1.54 | 1.13 | 0.84 | 3.42 |
| $W^{6+}$ exclusively on $B''$(f) site | 2.03 | 1.47 | 0.84 | 5.86 |
| $W^{6+}$ equal (50:50) on both $B''$(c) and $B''$(f) sites | 1.65 | 1.21 | 0.84 | 3.88 |

Table S8: $Ba_2CuTe_{0.7}W_{0.3}O_6$ ($x$ = 0.3) structure when site occupancy refined

| Space Group: $C2/m$, No. 12, 300 K |
| --- |
| $R_P$ = 1.84, $R_{wp}$ = 2.60, $R_{exp}$ = 0.81, $\chi^2$ = 10.50, var. 86 |
| $a$ = 10.2090(3) Å, $b$ = 5.71596(6) Å, $c$ = 10.08392(33) Å, $\beta$ = 107.9255(9)° |
| $Vol.$ = 559.876(13) Å$^3$ |

| Site | Wyckoff Position | x | y | z | Site fraction | Uiso |
|---|---|---|---|---|---|---|
| Ba1 | 4i | 0.12949(10) | 0 | 0.37951(7) | 1.0 | 0.00715(23) |
| Ba2 | 4i | 0.28386(11) | 0 | 0.85026(7) | 1.0 | 0.01104(25) |
| Te1 | 2a | 0 | 0 | 0 | 0.430(1) | 0.00808 |
| W1 | 2a | 0 | 0 | 0 | 0.570(1) | 0.00808 |
| Te2 | 2d | 0 | 0.5 | 0.5 | 0.970(1) | 0.00358 |
| W2 | 2d | 0 | 0.5 | 0.5 | 0.030(1) | 0.00358 |
| Cu1 | 4i | -0.09429(20) | 0.5 | 0.21528(15) | 1.0 | 0.01029(47) |
| O1 | 4i | 0.13200(11) | 0.5 | 0.40293(83) | 1.0 | 0.03453(276) |
| O2 | 8j | -0.10944(57) | 0.73097(83) | 0.36866(43) | 1.0 | 0.00071(132) |
| O3 | 4i | 0.31263(84) | 0.5 | 0.87764(76) | 1.0 | 0.00932(213) |
| O4 | 8j | 0.04436(76) | 0.75009(13) | 0.88975(56) | 1.0 | 0.01433(131) |

Table S9: Ba$_2$CuTe$_{0.8}$W$_{0.2}$O$_6$ ($x$ = 0.2) structure when site occupancy refined

| Space Group: $C2/m$, No. 12, 300 K | | | | | | |
|---|---|---|---|---|---|---|
| $R_P$ = 1.34, $R_{wp}$ = 1.75, $R_{exp}$ = 0.82, $\chi^2$ = 4.54, var. 86 | | | | | | |
| $a$ = 10.21742(26) Å, $b$ = 5.7173(4) Å, $c$ = 10.08907(25) Å, $\beta$ = 107.9346(7)° | | | | | | |
| Vol. = 560.768(10) Å$^3$ | | | | | | |
| Site | Wyckoff Position | $x$ | $y$ | $z$ | Site fraction | Uiso |
| Ba1 | 4i | 0.12946(7) | 0 | 0.37943(5) | 1.0 | 0.00750(17) |
| Ba2 | 4i | 0.28370(8) | 0 | 0.85019(5) | 1.0 | 0.01068(18) |
| Te1 | 2a | 0 | 0 | 0 | 0.618(1) | 0.00666 |
| W1 | 2a | 0 | 0 | 0 | 0.382(1) | 0.00666 |
| Te2 | 2d | 0 | 0.5 | 0.5 | 0.982(1) | 0.00396 |
| W2 | 2d | 0 | 0.5 | 0.5 | 0.018(1) | 0.00369 |
| Cu1 | 4i | -0.09399(14) | 0.5 | 0.2151(1) | 1.0 | 0.0094(3) |
| O1 | 4i | 0.1330(8) | 0.5 | 0.4024(6) | 1.0 | 0.02412(178) |
| O2 | 8j | -0.1077(4) | 0.7310(6) | 0.3687(3) | 1.0 | 0.00361(97) |
| O3 | 4i | 0.3145(6) | 0.5 | 0.8746(5) | 1.0 | 0.00652(149) |
| O4 | 8j | 0.04697(5) | 0.7538(9) | 0.8905(4) | 1.0 | 0.01459(93) |

Table S10: Ba$_2$CuTe$_{0.9}$W$_{0.1}$O$_6$ ($x$ = 0.1) structure when site occupancy refined

| Space Group: $C2/m$, No. 12, 300 K | | | | | | |
|---|---|---|---|---|---|---|
| $R_P$ = 1.13, $R_{wp}$ = 1.54, $R_{exp}$ = 0.4, $\chi^2$ = 3.39, var. 86 | | | | | | |
| $a$ = 10.2247(2) Å, $b$ = 5.72007(4) Å, $c$ = 10.09322(23) Å, $\beta$ = 107.9570(9)° | | | | | | |
| Vol. = 559.876(13) Å$^3$ | | | | | | |
| Site | Wyckoff Position | $x$ | $y$ | $z$ | Site fraction | Uiso |
| Ba1 | 4i | 0.12936(6) | 0 | 0.37930(4) | 1.0 | 0.00786(15) |
| Ba2 | 4i | 0.28331(7) | 0 | 0.85007(4) | 1.0 | 0.00984(16) |
| Te1 | 2a | 0 | 0 | 0 | 0.809(1) | 0.00567 |
| W1 | 2a | 0 | 0 | 0 | 0.191(1) | 0.00567 |
| Te2 | 2d | 0 | 0.5 | 0.5 | 0.991(1) | 0.00497 |
| W2 | 2d | 0 | 0.5 | 0.5 | 0.009(1) | 0.00497 |
| Cu1 | 4i | -0.09367(12) | 0.5 | 0.21510(9) | 1.0 | 0.00827(28) |
| O1 | 4i | 0.13281(62) | 0.5 | 0.40094(45) | 1.0 | 0.01419(141) |
| O2 | 8j | -0.10638(37) | 0.73184(56) | 0.36854(28) | 1.0 | 0.00675(89) |
| O3 | 4i | 0.31591(51) | 0.5 | 0.87345(44) | 1.0 | 0.00737(133) |
| O4 | 8j | 0.04899(49) | 0.75696(80) | 0.89113(35) | 1.0 | 0.01453(83) |

### 4. Extended X-ray Absorption Fine Structure (EXAFS) data for Ba$_2$CuTe$_{0.7}$W$_{0.3}$O$_6$

The site occupancy of W within the monclinic structre of Ba$_2$CuTe$_{0.7}$W$_{0.3}$O$_6$. was investigated by analysis of W $L_3$ edge EXAFS data. Models of the local environment of assumed: (1) full W$^{6+}$ substitution on the $B''$(c) site; and (2) full W$^{6+}$ substitution on the $B''$(f) site. As shown below, data were adequately fitted without inclusions of nearest neighbour W…Te paths, and, therefore, potential W…W scattering paths could be reasonably neglected.

Model (1) provides a plausible environment for $W^{6+}$ doping on the $B''$(c) site. As summarised in Table S11, all significant scattering paths within a radial distance of 4 Å from the W absorber are fitted, with reasonable contact distances and positive Debye-Waller factors. The EXAFS determined W-O bond length (1.900(7) Å) is close to the average Te/W-O bond length on the Te/W(1)$O_6$ site (1.920(5) Å) determined from the 300 K neutron diffraction data. It should be noted there is a slight variance in the Te/W(1)-O bond lengths (owing to the low symmetry) with the W/Te(1)-O(3) length being 1.902(2) Å and the W/Te(1)-O(4) being slightly longer at 1.930(2) Å. This difference is too small to be observed by EXAFS. The Bond Valence Sum determined for the W environment was 6.3(8) v.u.², consistent with the expected oxidation state. Fig. S15 compares shows the calculated model fit in comparison with the experimental data as $k^2\chi(k)$ and $\chi(R)$, its Fourier transform. The model clearly affords an excellent fit to the experimental data, as evident by the graphical fit and low $R$-factor of 1.18% (Table S11).

In contrast model (2), fails to provide a plausible description for $W^{6+}$ doping on the $B''$(f) site. As shown in Table S12, although model (2) affords a reasonable W(Te2)…O3 path length, the associated uncertainty is two orders of magnitude greater than determined for Model 1, and the Debyer-Waller factor is negative. Several other refined path lengths also have negative Debye-Waller factors, as shown in Table S12. The graphical comparison of the calculated model fit and experimental data, Fig. 17, shows noticeably poor agreement in several regions, which is reflected in the high $R$-factor of 9.07%. Consideration of the contribution of individual paths to the $\chi(R)$ data, show that the region 3<$R$<4 Å region is only well fitted by the W(Te1)…O4.1, W(Te1)…O4.1…Cu1.1, and W(Te1)…O4.1…Cu1.1…O4.1 paths afforded by $W^{6+}$ doping on the Te(1) site as shown in Fig. S16. In contrast, the local environment of $W^{6+}$ doping on the Te(2) site does not afford scattering paths to adequately fit the data in this range, as shown in Fig. S18.

The possibility of $W^{6+}$ substitution of the Cu was also examined, however the model failed to provide a physically credible solution. In particular, the $S_0^2$ parameter, which accounts for relaxation of the absorber atom in the presence of the core hole refined to a negative, and hence meaningless, value; for transmission measurements a value in the range 0.7 < $S_0^2$ < 1.0 is normally expected. In addition, refinement of some path lengths converged to implausible values or had associated Debye-Waller factors which were negative.

Our analysis of W $L_3$ EXAFS data therefore provides evidence for preferential W substitution on the $B''$(c) site, in agreement with the diffraction data, but has the advantage of providing an element specific perspective. Attempts were made to fit the data using contributions from both models (1) and (2), under linear restraints, to assess the potential for disorder of a fraction of $W^{6+}$ from the $B''$(c) to $B''$(f) site. However, it was not possible to adequately stabilise such a fit, since the number of variables approached the number of data points. Yet, there is some evidence for a contribtuion from the model (2) when comparing the $\chi(R)$ EXAFS plot in the range 2.0 < $R$ < 3.0. In Fig. S15, there is a slight intensity mismatch between the fit and data for the peaks in at a radial distance of ~2.1 and ~2.7 Å. In contrast, these same peaks are described quite well in Fig. S17 where the model (2) structure is used. Therefore, it is feasible this intensity mismatch in Fig. S15 between 2.0 < $R$ < 3.0 represents the contribution from the $B''$(f) site which were unable to stabilise in the fit given low 5% $W^{6+}$ site occupancy shown in the synchrotron X-ray data.

Table S11: Refined model parameters for W substitution on Te(1) site of $Ba_2CuTe_{0.7}W_{0.3}O_6$. Note that the path descriptions are those generated by the ATOMS algorithm to define the local cluster within Artemis.[3] $R$ is the refined path length of the specified path and $\sigma^2$ is the EXAFS Debye-Waller factor; $N$ is the path degeneracy, which corresponds to the co-ordination number in the first shell. The global parameters are: $S_0^2$, the passive electron reduction factor; the energy alignment factor, $\Delta E_0$; the number of independent data points, according to the Nyquist criterion, $N_{idp}$; and the number of variables in the model, $N_{var}$. The fitted ranges were 3.0 < $k$ < 11.0, with a Hanning window of $dk$ = 1.0 Å$^{-1}$; and 1.15 < $R$ < 4.0.

| Shell | Path | N | R(Å) | $\sigma^2$(Å$^2$) | Global parameters |
|---|---|---|---|---|---|
| 1 | W(Te1)…O3.1 (single scattering) | 6 | 1.900(7) | 0.002(1) | |
| 2 | W(Te1)…O3.1…O4.1 (double scattering) | 20 | 3.24(13) | 0.003(2) | |
| 2 | W(Te1)…Ba2.1 (single scattering) | 8 | 3.61(14) | 0.007(3) | $S_0^2$ = 0.76(6) $\Delta E_0$ = 7.3(1.0) eV $N_{idp}$ = 14 $N_{var}$ = 8 R(%) =1.18 |
| 2 | W(Te1)…O3.1 (hinge) | 24 | 3.80(15) | 0.001(1) | |
| 2 | W(Te1)…O4.1 (forward though absorber) | 4 | 3.82(15) | 0.002(5) | |
| 2 | W(Te1)…O4.1…Cu1.1 (forward scattering) | 8 | 3.85(15) | 0.008(2) | |
| 2 | W(Te1)…O4.1…Cu1.1…O4.1 (double forward scattering) | 4 | 3.85(15) | 0.006(3) | |

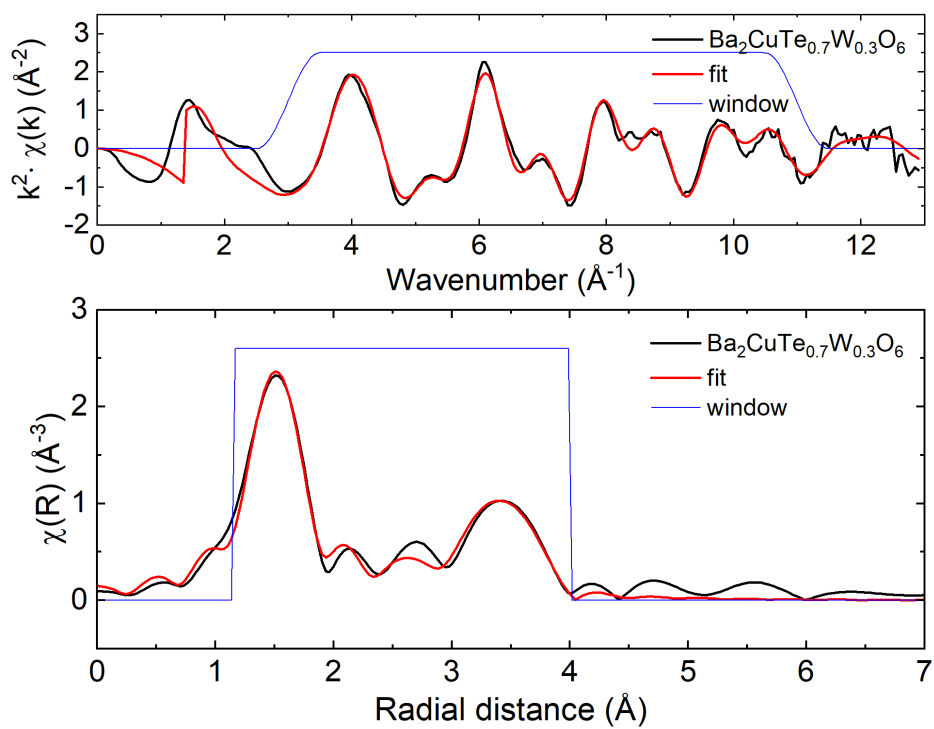

**Fig. S15** $k^2\chi(k)$ and $\chi(R)$ W $L_3$ EXAFS data of $Ba_2CuTe_{0.7}W_{0.3}O_6$ with model (1), assuming $W^{6+}$ doping on Te(1) site (uncorrected for phase shift). Solid black lines represent experimental data and red lines represent the model fits. Fitting windows are indicated by solid blue lines.

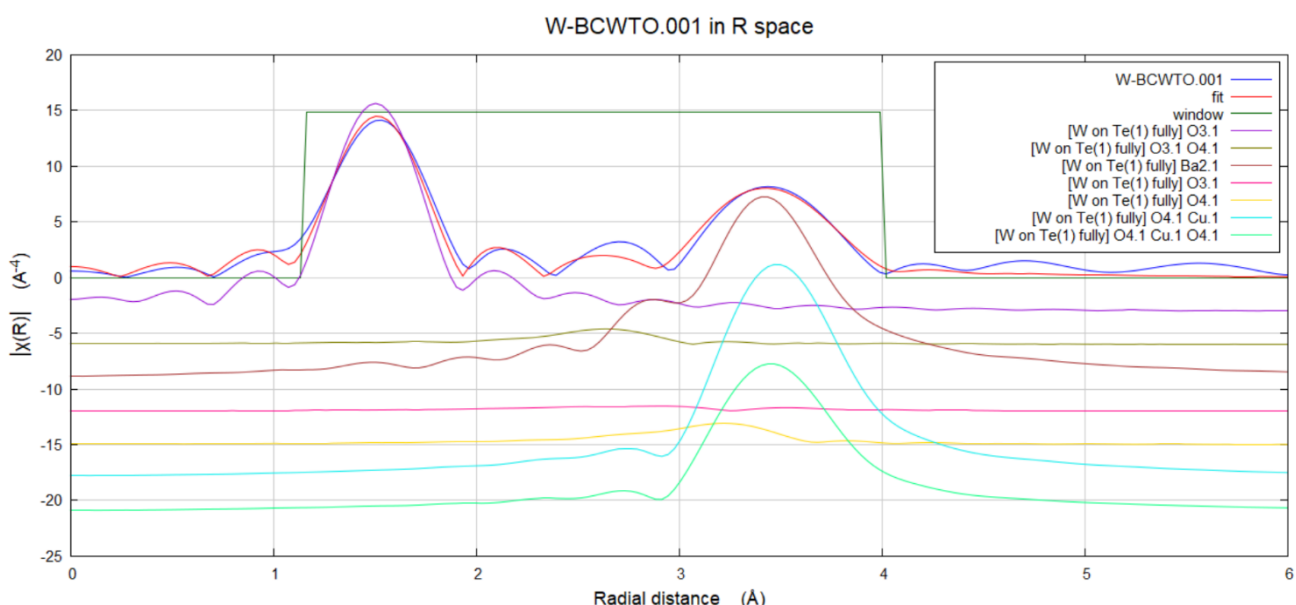

**Fig. S16** Showing component path contributions to $\chi(R)$ W $L_3$ EXAFS data of Ba$_2$CuTe$_{0.7}$W$_{0.3}$O$_6$ with model (1), assuming W substiution on Te(1) site (uncorrected for phase shift). Solid blue line represents experimental data and red line represents the model fit; other solid lines represent individual path contribution summarised in Table S11. The fitting window is shown indicated by the solid green line.

Table S12: Refined model parameters for W substitution on Te(2) site. Note that the path descriptions are those generated by the ATOMS algorithm to define the local cluster within Artemis.[3] $R$ is the refined path length of the specified path and $\sigma^2$ is the EXAFS Debye-Waller factor; $N$ is the path degeneracy, which corresponds to the co-ordination number in the first shell. The global parameters are: $S_0^2$, the passive electron reduction factor; the energy alignment factor, $\Delta E_0$; the number of independent data points, according to the Nyquist criterion, $N_{idp}$; and the number of variables in the model, $N_{var}$. The fitted ranges were 3.0 < $k$ < 11.0, with a Hanning window of $dk$ = 1.0 Å$^{-1}$; and 1.15 < $R$ < 4.5.

| Shell | Path | $N$ | $R$(Å) | $\sigma^2$(Å$^2$) | Global parameters |
|---|---|---|---|---|---|
| 1 | W(Te1)…O1.1 (single scattering) | 6 | 1.90(15) | -0.0004(25) | $S_0^2$ = 0.62(13) $\Delta E_0$ = 7.6(2.3) eV $N_{idp}$ = 14 $N_{var}$ = 8 $R$(%) =9.07 |
| 2 | W(Te1)…Cu1.1 (double scattering) | 2 | 2.69(21) | 0.03(10) | |
| 2 | W(Te1)…O1.1…O2.1 (double scattering) | 16 | 3.24(26) | -0.006(12) | |
| 2 | W(Te1)…Ba1.1 (single scattering) | 4 | 3.45(27) | 0.03(0.16) | |
| 2 | W(Te1)…O1.1 (forward though absorber) | 6 | 3.80(30) | -0.012(3) | |
| 2 | W(Te1)…O2.1…Ba2.2 (non-forward linear) | 8 | 4.69(37) | 0.008(0.012) | |

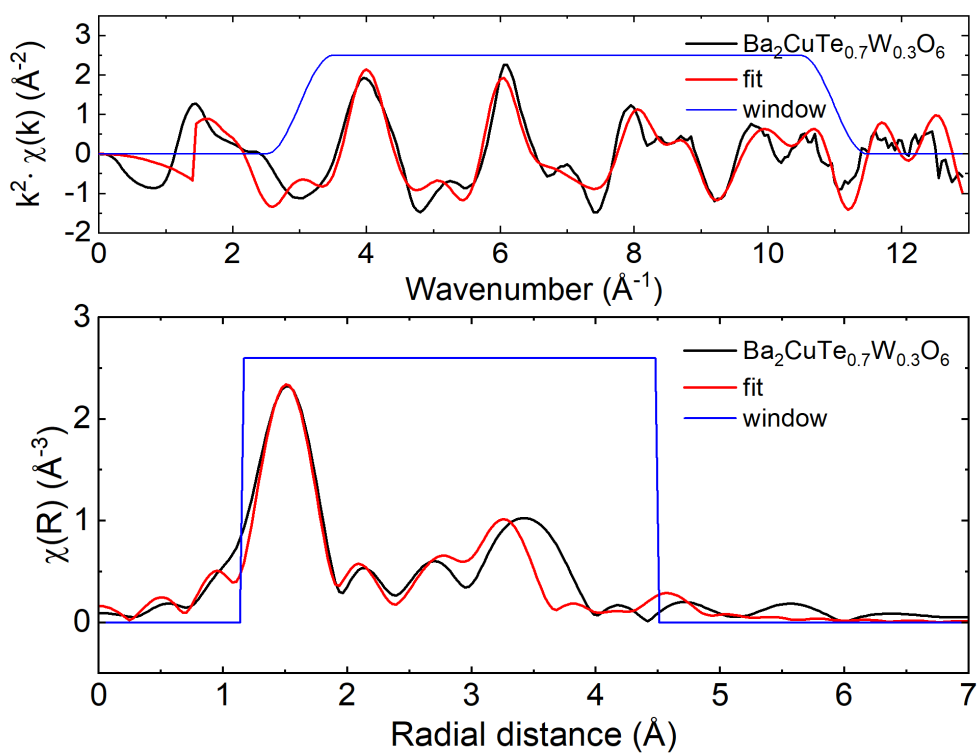

**Fig. S17** $k^2\chi(k)$ and $\chi(R)$ W $L_3$ EXAFS data of $Ba_2CuTe_{0.7}W_{0.3}O_6$ with model (2), assuming $W^{6+}$ doping on Te(2) site (uncorrected for phase shift). Solid black lines represent experimental data and red lines represent the model fits. Fitting windows are indicated by solid blue lines.

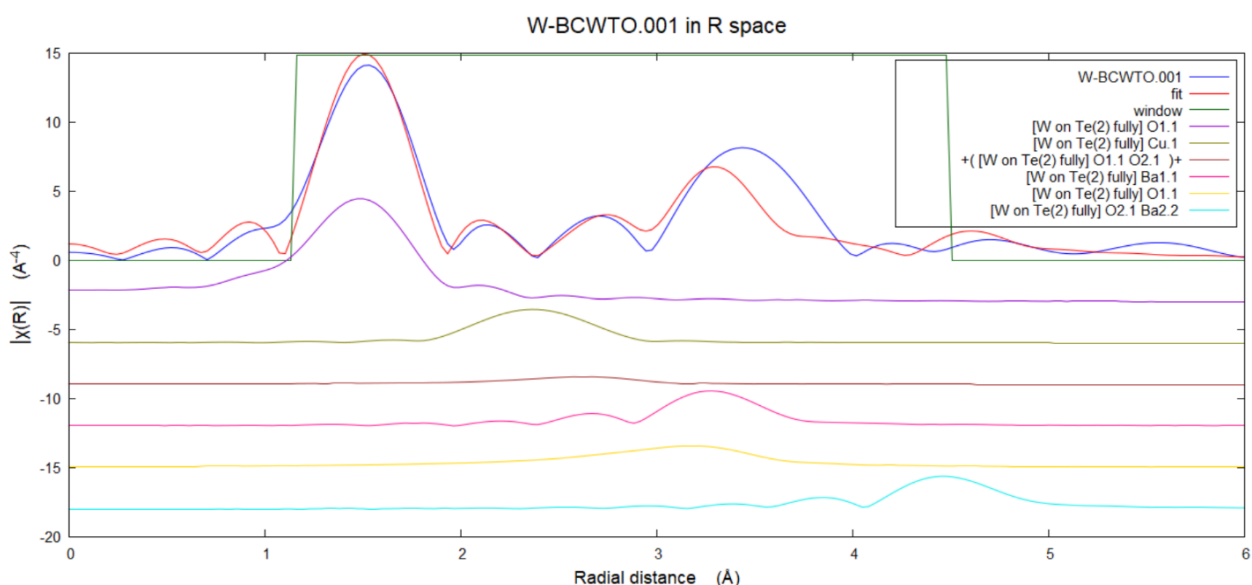

**Fig. S18** Showing component path contributions to $\chi(R)$ W $L_3$ EXAFS data of $Ba_2CuTe_{0.7}W_{0.3}O_6$ with model (2), assuming $W^{6+}$ doping on Te(2) site (uncorrected for phase shift). Solid blue line represents experimental data and red line represents the model fit; other solid lines represent individual path contribution summarised in Table S12. The Fitting window is shown indicated by the solid green line.

## 5. Magnetic Susceptibility

All the data were fitted to the inverse Curie-Weiss law $1/\chi = (T-\theta_W)/C$. Below is an example Curie-Weiss fit for $Ba_2CuTe_{0.9}W_{0.1}O_6$.

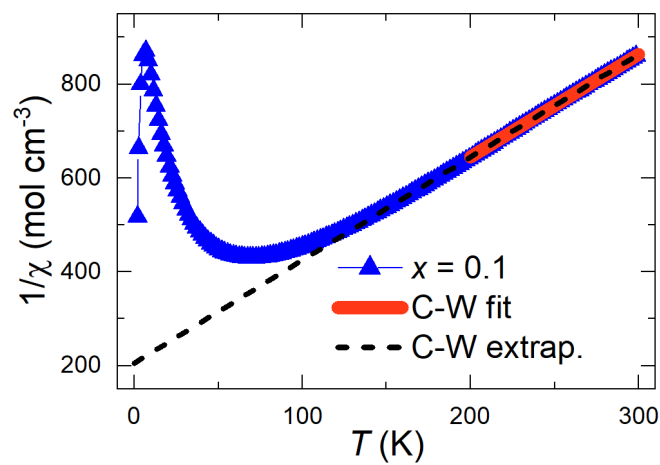

**Fig. S19** Curie-Weiss fit the inverse $Ba_2CuTe_{0.9}W_{0.1}O_6$ susceptibility data.

The Cu-O(1-4)-Te(1,2) bond angles and average Cu-O bond lengths determined from the synchrotron X-ray data for $x$ = 0.1, 0.2 and 0.3 were compared to the corresponding Weiss constant ($\theta_W$) for each sample. The plots in Fig. S20 show there is little change in the average Cu-O bond length with increasing $W^{6+}$ content. The largest change is in the Cu-O(3)-$B''$(c) bond angle which increases by nearly 4° between $x$ = 0.1 and $x$ = 0.3. There is a smaller change in the Cu-O(4)-$B''$(c) angle and the Cu-O(1,2)-$B''$(f) bond angles change even less. In line with Iwanaga et al. (1999), in increase in the Cu-O(3)-$B''$(c) corresponds with a decrease in the value of $\theta_W$ as $x$ increases, implying a similar correlation as to that observed in double perovskite $(Ba/Sr)_2Cu(Te/W)O_6$ structures.[4]

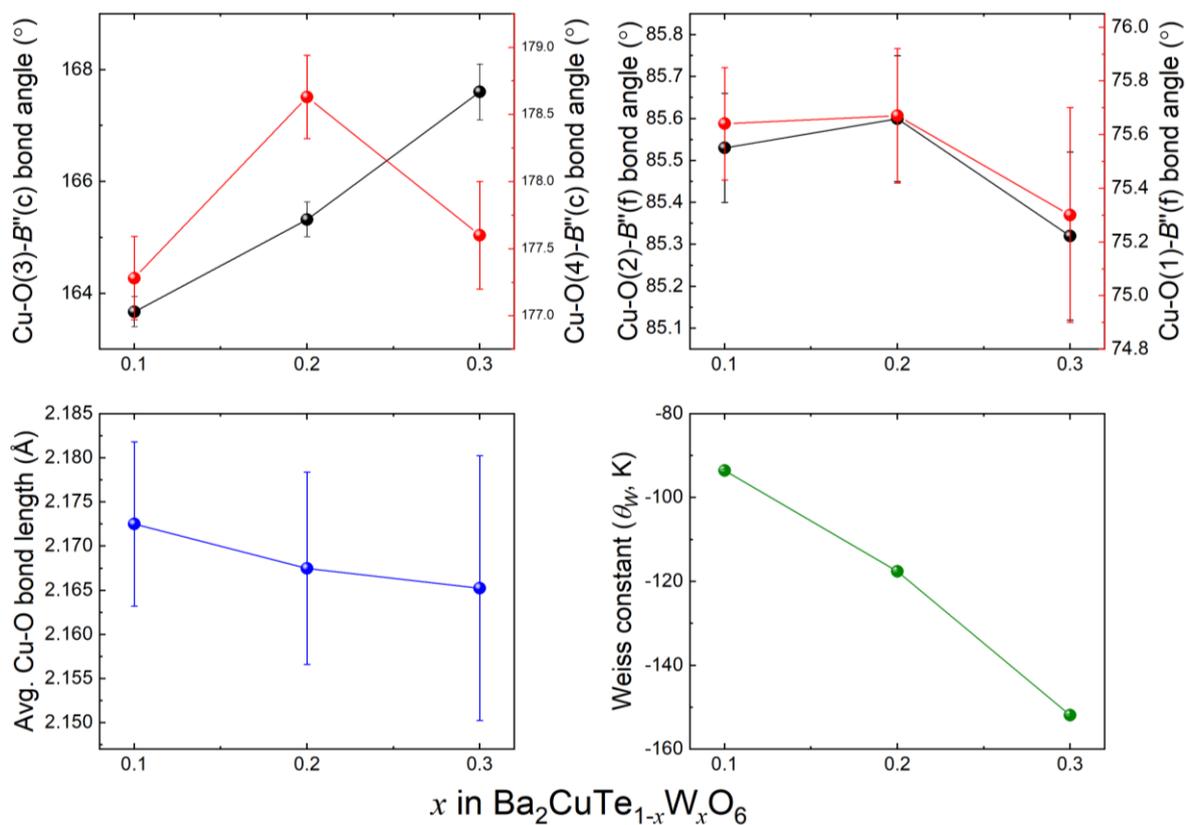

**Fig. S20** Comparisons of the Cu-O bond lengths, Cu-O-$B''$(c) and Cu-O-$B''$(f) bond angles vs the Weiss constants for the $x$ = 0.1, 0.2 and 0.3 samples.